\documentclass[aps,pre,twocolumn,showpacs,groupedaddress,footinbib]{revtex4-1}
\usepackage{natbib,hyperref}
\usepackage[english]{babel}
\usepackage[utf8]{inputenc}
\usepackage{graphicx}
\usepackage[caption=false]{subfig}
\usepackage{amsmath}
\usepackage{amsfonts}
\usepackage{amssymb}
\usepackage{color,soul}
\setulcolor{red}
\usepackage{xcolor}

\begin{document}

\title{Phase separation in a two-dimensional binary colloidal mixture by quorum sensing activity}

\author{Jalim Singh}
\author{A. V. Anil Kumar}
\email{anil@niser.ac.in}
\affiliation{School of Physical Sciences, 
National Institute of Science Education and Research,
HBNI, Jatni, Bhubaneswar 752050, India.}

\begin{abstract}

	We present results from Langevin dynamics simulations of a glassy 
	active-passive mixture of soft-repulsive binary colloidal disks. Activity
	on the smaller particles is applied according to the quorum sensing
	scheme, in which a smaller particle will be active for a persistence time 
	if its local nearest neighbors are equal to or greater than a certain 
	threshold value. We start with a passive glassy state of the system and 
	apply activity to the smaller particles, which shows a non-monotonous glassy
	character of the active particles with the persistence time of the active
	force, from its passive limit (zero activity). On the other hand, passive 
	particles of the active-passive mixture phase separate at the intermediate 
	persistence time of the active force, resulting in the hexatic-liquid and
	solid-liquid phases. Thus, our system shows three regimes as active glass,
	phase separation, and active liquid, as the persistence time increases from
	its smaller values. We show that the solidlike and hexatic phases consisting of
	passive large particles are stable due to the smaller momentum transfer from 
	active to passive particles, compared to the higher persistence time where the
	positional and orientational ordering vanishes. Our model is relevant to
	active biological systems, where glassy dynamics is present, e.g., 
	bacterial cytoplasm, biological tissues, dense quorum sensing bacteria, and 
	synthetic smart amorphous glasses.

\end{abstract}

\maketitle

\section{Introduction}
Active (self-propelled) particles consume energy from within the system or 
by some external means, and drive themselves to a non-equilibrium
state \cite{j:ramaswamy_activI,j:bechinger_rev_activ}. Systems with self-propelled 
particles are of great interest from the perspective of fundamental physics and
they are ubiquitous. Examples of living active matter systems are, but 
not limited to, sperm swarming \cite{j:sperm_swarm}, bird-flocking and fish 
schools \cite{b:group_living}, biological microswimmers \cite{j:berg_chemotaxis},
quorum sensing bacteria \cite{j:qs_bassler} etc. Another class of active 
systems is synthetic active matter; examples include artificial
microswimmers \cite{j:bechinger_rev_activ}, active mechanical (micro and nano) 
robots \cite{j:robots}, and synthetic quorum sensing systems \cite{j:qs_bechinger} 
that are useful for the targeted drug delivery, directed mechanical work, 
biomarkers, and local-density-dependent motility. Motile bacteria sense a
local concentration of the signaling molecules and themselves (e.g., 
acyl-homoserine lactones in Gram-negative bacteria) to perform virulence, biofilm 
formation, etc., which is termed as quorum sensing (QS) \cite{j:qs_review}.
Bacteria use QS to produce, release, and sense the extracellular chemical 
signals that are called as autoinducer molecules for cell-cell communication.
The number of autoinducer molecules increases as a function of the bacterial cell
density, which activate its gene transcriptions at a cell density threshold:
bacteria perform QS at both low and high cell density 
thresholds \cite{j:lowhighQS}. 

Several living systems including bacterial cytoplasm and collective cell migration,
show the fingerprints of glassy dynamics \cite{j:bactcytoplsmglas,j:actglas_rev}.
Glasses are dense amorphous systems, and one of the hallmarks of the glassy systems
is a dramatic slowdown of the density relaxations with a minimal change in 
their structure \cite{j:berthier}. The glassy dynamics of the bacterial
cytoplasm arises from its crowded (dense) intracellular components, which 
fluidizes by the metabolic activities of the bacterial cell, showing an active
glassy characteristic \cite{j:bactcytoplsmglas}. Another class of active-glass
systems are smart amorphous materials (artificial systems) including
phase-changing materials and self-healing glasses \cite{j:actglas_rev}. Experimental
study of these active materials is difficult at high density, where the corresponding 
passive system shows glassiness. Very recently, Klongvessa \textit{et al.} performed 
an experiment on the gold colloidal particles that are called as active Janus colloids,
half coated with platinum at high volume fractions \cite{j:januscolid_exp1}. The
authors concluded that active glass slows down at the smaller activity, whereas
its fluidization enhances at the higher activity, which is a non-monotonous character of
the active glassy Janus colloids. Simulation studies by Szamel and 
Berthier \cite{j:szamel_activ,j:szamel_activ_rev} also show the non-monotonous
character in the (all-)active glassy binary colloidal mixture. Another simulation
study of the active-passive binary mixture shows that activity fluidizes the 
glassy state \cite{j:chandan_actbin}.

Active DNAs in the interphase of chromosome drive to the segregated domains, consisting
of euchromatin and heterochromatin that is explained from the phase separation of 
active-passive components \cite{j:chromatin}. Recently, a study of activity-induced
phase separation in the monodisperse soft-repulsive active-passive disks by
Stenhammer \textit{et al.} \cite{j:cates_act_phasep} showed the presence of 
segregated active and passive domains, where the compression waves originating from
the corona of active particles causes the crystallization of the passive particles.
In a soft-repulsive active dumbbells system, the liquid (or gas) and hexatic phase
coexist, as reported by Cugliandolo \textit{et al.} \cite{j:suma_phascoexist}.
These studies also showed that the motility-induced phase 
separation (see Ref. \cite{j:mips_flock} and references therein), where phase 
separation occurs at a threshold value of the activity, which is not true in general.
Thus, it prompts us to study the activity-induced phase separation in a dense 
active-passive mixture. In this study, we simulate a two-dimensional (2D) binary 
colloidal mixture consisting of small and large size particles in its glassy state.
The small particles are kept motile depending upon their local nearest
neighbors (local density), $n_b^{fcs}$, to replicate the local density-dependent
sensing in QS bacteria. Thus, our
system could be a model system for the smart amorphous materials, and the QS in
crowded bacteria. We use three control parameters in our simulations: activity $f_a$,
persistence time $\tau_p$, and the $n_b^{fcs}$ that determines the number density of
the active particles in our system. Starting from the passive glass, the active system
shows enhanced glassiness at smaller $\tau_p$ (mixed-phase), which first phase separate
into hexatic-liquid and then to the solid-liquid at the intermediate $\tau_p$; solidlike
and hexatic phases in the phase separation regime are consist of only the
large (passive) particles. On further increment in $\tau_p$, activity $f_a$ fluidizes
both species of the particles that undergo mixing. Thus, our system shows a
continuous increase in hexatic order from its passive limit, which becomes solidlike
at the intermediate $\tau_p$. This solidlike phase melts into the hexatic phase, 
which subsequently becomes liquid at the higher $\tau_p$.

\section{Model and simulation details}
In this section, we discuss Langevin dynamics simulations, the QS activity,
and an initial configuration of our 2D binary colloidal system.

\subsection{Equations of motion \label{s:em}}
We simulate a 50:50 mixture, consisting of 1000 soft colloidal disks of size 
ratio 1:1.4 that has sufficient frustration to prevent crystallization, and is a well-known
glass-forming system \cite{j:onuki_2dglass,j:tanaka_2dglasorder}. Equations of motion for
the colloidal disks are of stochastic type, i.e., Langevin equations 
\begin{equation}
	\label{e:lde}
	m_i \ddot{\mathbf r_i} = -\gamma \dot{\mathbf r_i} + \sum_{ij}
	\mathbf F_{ij} + \mathbf F^a_i + {\sqrt{2k_BT\gamma}} {\boldsymbol {\eta}_i} , 
\end{equation}
where $\gamma$ is the friction coefficient, 
$\mathbf F_{ij}=- \boldsymbol{\nabla} V(r_{ij})$ is the inter-particle interaction
force, $T$ is the thermal noise temperature, $\mathbf F^a_i$ is an active force of 
magnitude $f_a$, and $\eta_i$ is the random gaussian noise with zero mean and unit 
variance, i.e., $\langle \eta_i \rangle =$0 and 
$\langle \eta_{i\alpha}(t)\eta_{j\beta}(t^{\prime}) \rangle = 
\delta_{ij}\delta_{\alpha\beta}\delta(t-t^{\prime})$.
Inter-particle interactions are modelled by purely repulsive and soft Lennard-Jones 
potential \cite{ham} 
\begin{equation}
	V(r_{ij}) = 4\epsilon\left[\left(\frac{\sigma_{\alpha\beta}}{r_{ij}}\right)^{12} 
	- \left(\frac{\sigma_{\alpha\beta}}{r_{ij}}\right)^{6} + \frac{1}{4}\right],
\end{equation}
with cut-off at $r^{\alpha\beta}_c=2^{1/6}\sigma_{\alpha\beta}$.
Here $(\alpha,\beta) \in (A,B)$, $\epsilon=$ 1.0, $\sigma_{AA} =$ 1.4 $\sigma_{BB}$, 
$\sigma_{AB}=$ 1.2 $\sigma_{BB}$, and $\sigma_{BB} =$ 1.0; distances are measured in the
units of $\sigma_{BB}$. For the underdamped case, we use $\gamma=$ 10.0. The system is 
simulated at a constant area fraction $\phi=$ 0.628 that is calculated
as $\phi=\pi\rho(\sigma^2_{AA}+\sigma^2_{BB})/8$.  The equations of motion are integrated 
using a second-order scheme given by Vanden-Eijnden and Ciccotti \cite{j:eijden-ciccotti}, 
using the time step $dt=$ 0.002, for active and corresponding passive system. Note that 
the active system will become passive in the limit of activity $f_a=$ 0 (zero active force). 
All the quantities presented here are in the Lennard-Jones (reduced) units, i.e., 
reduced density $\rho^*=	\rho\sigma^3$, reduced temperature $T^*=k_BT/\epsilon$, reduced
time $t^*=(\epsilon/{m\sigma^2})^{1/2}t$, and reduced force $f^*=f\sigma/\epsilon$ \cite{ham}. 

\subsection{Quorum sensing activity \label{s:qs}}
\begin{figure}
	\includegraphics[width=8.5cm, height=5.3cm]{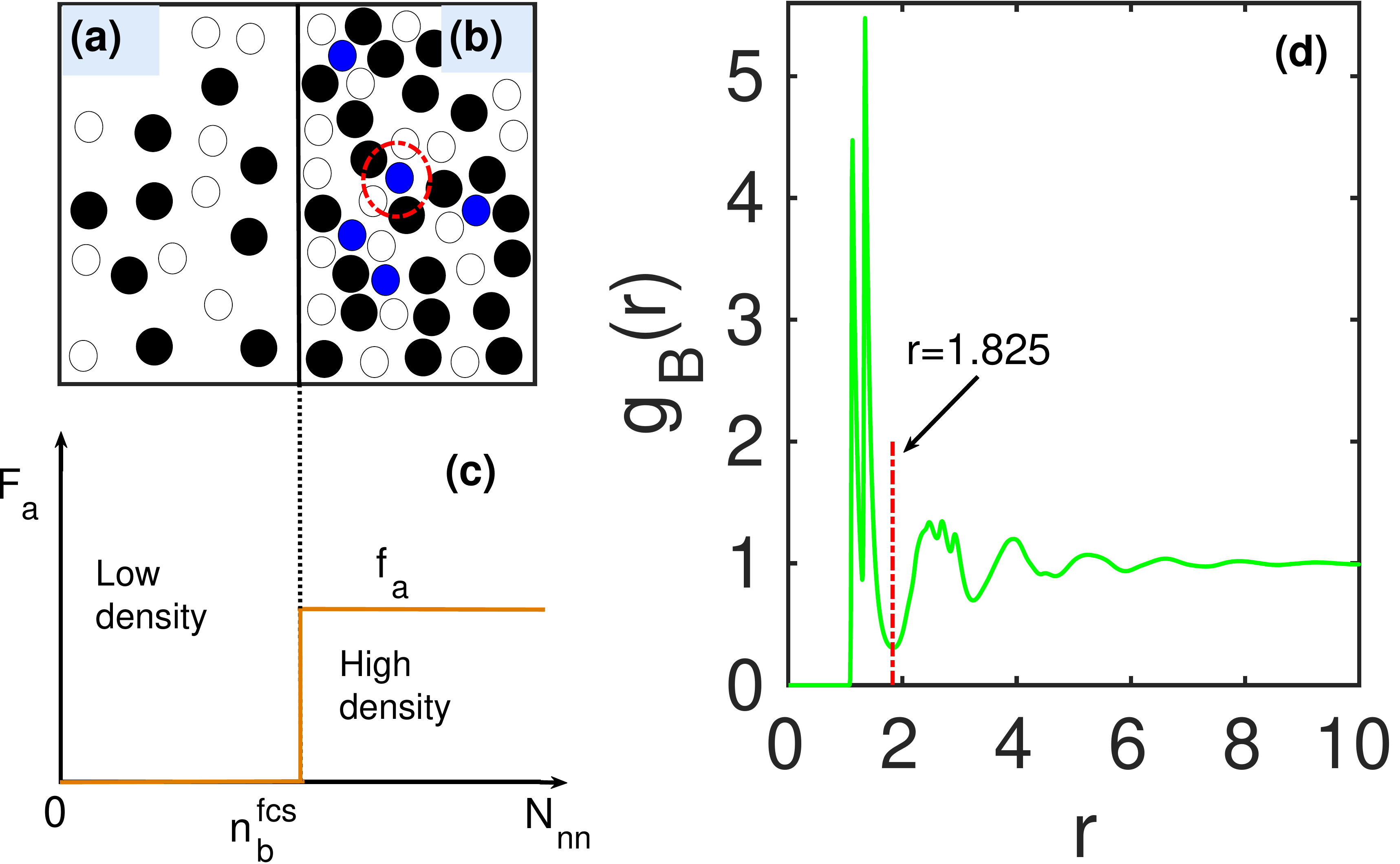}
	\caption{\label{f:qshd}(a--c) Schematic diagram for the quorum sensing: (a) Passive 
	binary mixture, (b) active-passive binary mixture, (c) quorum sensing rule at high
	density. In (a) and (b), open small circles are passive $B$ (small) particles, whereas
	filled black circles are passive $A$ (large) particles. In (b) filled blue circles are
	the active $B$ (small) particles and red dot-dashed circle represents the FCS of one 
	of the active particles. (d) RDF of $B$ particles. The red dot-dashed line at $r=$ 1.825
	corresponds to the FCS in $g_B(r)$.}
\end{figure}

QS bacteria uses local (neighboring) density detection algorithm to express their genes,
virulence, biofilm formation etc. \cite{j:qs_bassler,j:qs_review,j:qs_bechinger}. 
To replicate the behavior of QS bacteria, we simulate the binary colloidal mixture in two
dimensions in which only small particles are active according to the local density 
searching algorithm for a finite persistence time of the active force. Note that all
$A$ (large) particles are passive throughout the simulations. For the range of local
density sensing algorithm, we compute radial distribution function (RDF) of $B$ particles
for the passive system at $T=$ 0.01 and $\phi=$ 0.628 as
\begin{equation}
	g_B(r) = \frac{A}{2\pi r \Delta r N N_B} \left\langle \sum_{i=1}^{N_B}
	\sum_{\substack{j=1 \\j \ne i}}^N \delta({r - |\mathbf r_j - \mathbf r_i|}) \right\rangle ,
\end{equation}
which is shown in Fig. \ref{f:qshd}(d). Here, $A$ is the area of the two-dimensional 
simulation box, $N$ and $N_B$ are total number of particles and number of $B$ particles
in the system, respectively. The $g_B(r)$ computes number of $B$ and $A$ particles around a
tagged $B$ particle, its first peak corresponds to the $B-B$ particles,
whereas the second peak corresponds to the $B-A$ particles. A minima at $r=$ 1.825, after
the second peak, shown by red dot-dashed line in Fig. \ref{f:qshd}(d), corresponds to the
first coordination shell (FCS) of $g_B(r)$. We compute number of nearest-neighbors
(both $A$ and $B$) of each $B$ particle within the FCS. A small ($B$) particle is active
according to a criterion that if its $N_{nn}^i>=n^{fcs}_b$. Here, $N_{nn}^i$ is the 
instantaneous number of nearest neighbors of a $B$ particle within the 
FCS [see Figs. \ref{f:qshd}(b) and \ref{f:qshd}(d)] and the $n^{fcs}_b$ is a fixed 
threshold value for
the number of nearest neighbors of $B$ particles. We scan $N_{nn}^i$ of each $B$ particle 
at each $t=\tau_p$, where $\tau_p$ is the persistence time of the active force. In this
study, we present results from the $n^{fcs}_b=$ 4 and 5. At $n^{fcs}_b=$ 6, only $10-15\%$ 
of the $B$ particles are 
active (data not presented in this study, which will be reported elsewhere), and 
at $n^{fcs}_b=$ 7 only $1-5$ $B$ particles are active. Thus, the number density of 
active $B$ particles, $\rho_{ab}$, is controlled by the QS scheme; the range 
of $\rho_{ab}=0.256-0.27$ for $n_b^{fcs}=$ 4 and $\rho_{ab}=0.2-0.265$ for $n_b^{fcs}=$ 5.
The shift in the range of $\rho_{ab}$ towards its smaller values is natural 
for $n_b^{fcs}=$ 5, as the number of $B$ particles with $N_{nn}^i=$ 5 or greater decrease
compared to the $n_b^{fcs}=$ 4, at the same volume fraction. Thus, the number density of 
active particles in our study is varying and according to the QS rules using the local 
neighbor searching algorithm.

We apply active force on $B$ particles according the QS scheme described above for the
finite persistence time $\tau_p$ in the range $0.01-20.0$. The active 
force $\mathbf F_a^i$ [see Eq. \ref{e:lde}] on an $i$th particle is given as 
$\mathbf F_a^i = f_a \boldsymbol \zeta_i$, where $f_a$ is a magnitude of the active force.
Figure \ref{f:qshd}(c) shows that if a $B$ particle satisfies the criterion mentioned above then
it is active with the activity $f_a$, otherwise it is passive and $f_a=$ 0. The direction of
the active force on an $i^{th}$ particle is given
by $\boldsymbol \zeta_i = \zeta_{ix}\hat{\boldsymbol x} + \zeta_{iy}\hat{\boldsymbol y}$ as 
\begin{equation}
	\langle\zeta_{i\alpha}(t)\zeta_{j\beta}(t^\prime)\rangle = 
	\delta_{ij}\delta_{\alpha\beta},
\end{equation}
for $|t-t^\prime|<=\tau_p$ and zero otherwise; $(\alpha,\beta) \in (x,y)$.
The $\zeta_{ix}$ and $\zeta_{iy}$ are between $-1$ and $+1$, chosen from the uniform random
deviates.

\subsection{Initial configurations}

\begin{figure}
	\includegraphics[width=8.5cm, height=4.0cm]{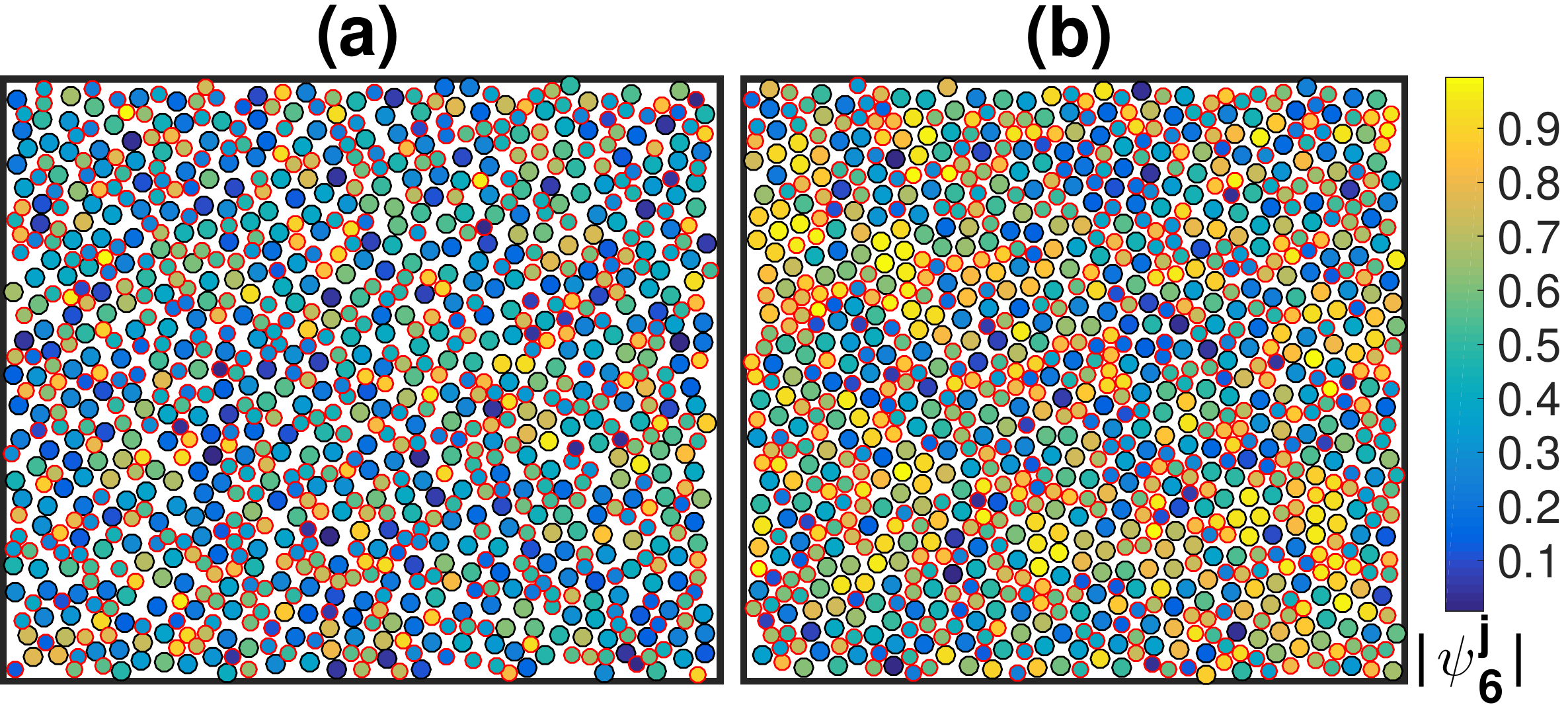}
	\caption{\label{f:cnf} (a) Random initial configuration of the Passive mixture at
	$\phi=$ 0.628, (b) Final configuration of the Passive mixture at $T=$ 0.01 and 
	$\phi=$ 0.628. Black (large) and red (small) edge spheres correspond to the $A$ and $B$ 
	particles. Face color of the particles is according to the color bar for local hexatic 
	order parameter, $|\psi_6^j|$.}
\end{figure}

We prepare a random initial configuration of the passive binary colloidal mixture at
the area fraction $\phi=$ 0.628, which is shown in Fig. \ref{f:cnf}(a). This random initial
configuration is equilibrated at temperature $T=$ 8.0 using the Langevin dynamics 
simulations. Final configuration of the equilibrated system is quenched from $T=$ 8.0
to $T=$ 0.01, which is equilibrated for $5\times 10^7$ steps, and then trajectories are
stored up to $2.5\times 10^6$ steps. The equilibrated configuration of the passive system 
at $T=$ 0.01, is used as an initial configuration for the active system. Thus, the
active-passive binary mixture at all activities and persistence times runs for 
the $5\times 10^7$ steps to reach its steady state, after that, trajectories of the system
are stored for analysis purpose. To examine local orientational ordering in the passive
system, we invoke single particle hexatic order parameter (HOP) that entails information
about the local hexatic order in two dimensions. The HOP is defined as 
\begin{equation}
	\label{e:psi6}
	\psi_6^j = \frac{1}{N_b^j} \sum_{m=1}^{N_b^j} \exp(\imath 6\theta_j^m) ,
\end{equation}
where $N_b^j$ are the number of nearest neighbors of a particle $j$, $\theta_j^m$ is the
angle between the radius vector $\mathbf r_j^m$ and the reference
axis \cite{j:hardisk_KTHNY,j:onuki_2dglass}. Depending upon the $\theta_j^m$, values of
the $|\psi_6^j|$ ranges from 0 to 1; a perfect trigonal corresponds to the $6\theta_j^m=2\pi$ that 
has a value $|\psi_6^j|=$ 1, whereas $|\psi_6^j|=$ 0 corresponds to a random orientational 
arrangement of six nearest-neighbors of a particle $j$. We use cut-off distances $2.085$ 
and $1.825$, respectively as the second minima of $g_A(r)$ and $g_B(r)$, for calculating 
nearest-neighbors of $A$ and $B$ particles; for reference the $g_B(r)$ is 
shown in the Fig. \ref{f:qshd}(d). The absolute values of $\psi_6^j$ of majority of particles 
for the random initial configuration (see Fig. \ref{f:cnf}(a)), are far less than 1.0, showing 
a random orientational order, except for few particles, for them it shows local hexatic 
ordering ($|\psi_6^j|\approx$ 1) because the natural number of nearest neighbors of a particle 
are six in a dense 2D system. The equilibrated final configuration of the passive system 
at $T=$ 0.01 shows the local hexatic ordering of $A$ and $B$ particles that can be seen
in Fig. \ref{f:cnf}(b). Fig. \ref{f:cnf}(b) shows that the local hexatic ordering increases
in the passive system at $T=$ 0.01 and the system shows mixing. Structural analysis of the 
passive system is given in Appendix \ref{s:glas_pasiv}, which shows that it is amorphous and 
does not posses long-range (or quasi-long-range) positional or orientational order.

\section{Results and discussion}

\subsection{Configurations of active-passive binary mixture}
\begin{figure}
	\includegraphics[width=9.0cm, height=7.5cm]{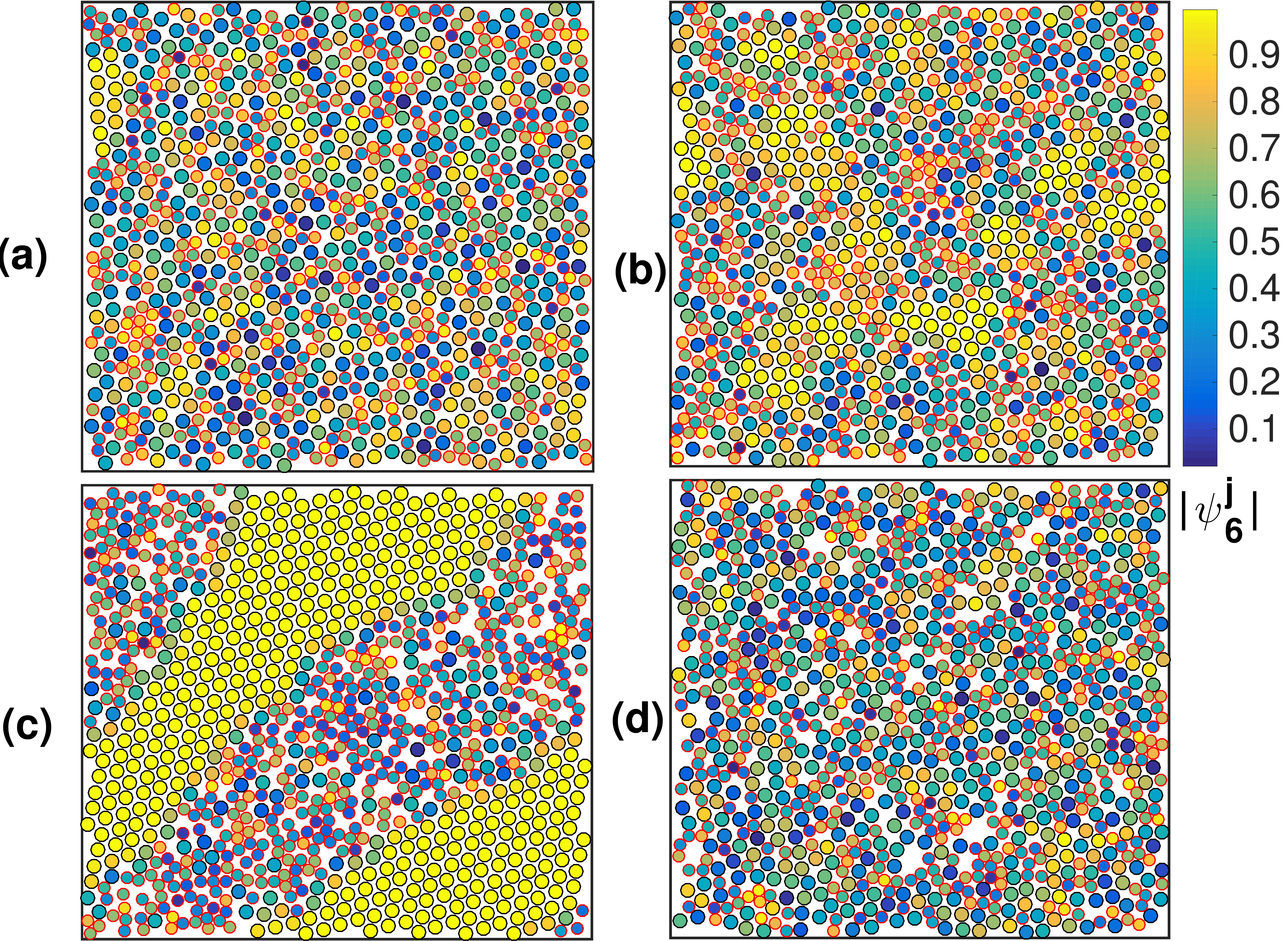}
	\caption{\label{f:hexpop} Steady-state configurations of the active-passive binary mixture 
	at $n_b^{fcs}=$ 4 and $f_a=$ 3.0 at persistence times $\tau_p=$ 0.1 (a), $\tau_p=$ 0.3 (b),
	$\tau_p=$ 3.0 (c), and $\tau_p=$ 15.0 (d). Black (large) and red (small) edge spheres 
	correspond to the $A$ and $B$ particles. Face color of the particles is according to the 
	color bar for local hexatic order parameter, $|\psi_6^j|$.}
\end{figure}

The final configuration obtained at $T=$ 0.01, shown in Fig. \ref{f:cnf}(b), is used 
as an initial configuration for the active system ($f_a\ne$ 0). As discussed in 
Sec. \ref{s:qs} that we apply activity to only smaller particles according to the QS 
rules. We vary the activity and the persistence time of the active force as $f_a=$1-10 
and $\tau_p=$ 0.01-20.0, respectively, for this study. In Fig. \ref{f:hexpop}, we show 
few typical configurations of the active-passive binary mixture at $n_b^{fcs}=$ 4 and 
$f_a=$ 3. Figure \ref{f:hexpop}(a) shows a configuration of the active-passive mixture at
smaller persistence time $\tau_p=$ 0.1. By comparing the local hexatic order $|\psi_6^j|$
of the particles in Fig. \ref{f:cnf}(b) (passive mixture) and 
Fig. \ref{f:hexpop}(a), we show that local hexatic order is similar in active-passive
mixture at smaller $\tau_p$ to the corresponding (all particles) passive system. However,
at $\tau_p=$ 0.3 [see Fig. \ref{f:hexpop}(b)], the local hexatic order increases compare
to the $\tau_p=$ 0.1, and the demixing of the large particles starts in the active-passive 
mixture. On further increasing $\tau_p$, system shows an extent of demixing of $A$ and $B$ 
particles [see Fig. \ref{f:hexpop}(c)], where almost all $A$ particles show hexatic order
parameter, $|\psi_6^j|\approx$ 1. On the other hand, $B$ particles that are active and
passive, show hexatic and random orientation ($|\psi_6^j|$ near 0); a majority of them
show random orientation [see Fig. \ref{f:hexpop}(c)]. Note that we do not distinguish active
and passive \textit{B} particles, at this point. At $\tau_p=$ 15, both $A$ and $B$ particles
show mixing again and the $|\psi_6^j|$ of the particles decrease, even smaller than that at the
smaller persistence times [see Fig. \ref{f:hexpop}(d)]. Thus, our active-passive mixture shows
phase separation at the intermediate $\tau_p$, where hexatic order increases consisting of $A$ 
particles; the hexatic order decreases at the higher $\tau_p$. Thus, the hexatic order in $A$
particles varies non-monotonically with $\tau_p$. However, the $B$ particles that
are active and (few of them) passive both, show decrement in the hexatic order in the phase 
separation regime.  

\subsection{Phase separation observables\label{s:pso}}

The configurations of the active-passive mixture show that the system phase separates
at the intermediate $\tau_p$ and the hexatic order increases consisting of large particles.
To quantify the phase separation, we divide the simulation box into a number of square 
cells ($N_{cell}=8\times8$) and calculate the difference in the number of $A$ and $B$ 
particles, which can be obtained by choosing the cell length,
$l_c=$ 5.38 (along each spatial direction of the box length, $L$), for this study. 
The difference in the number of $A$ and $B$ particles is calculated as
\begin{equation}
	\label{e:psop}
	\Phi(f_a,\tau_p) = \frac{1}{N_{cell}} \left\langle \sum_{i=1}^{N_{cell}}
	\frac{|n_A^i-n_B^i|}{(n_A^i+n_B^i)} \right\rangle,
\end{equation}
where $n_A^i$, $n_B^i$ are the number of $A$ and $B$ particles in an $i$th cell,
respectively \cite{j:activ_phassep_chandan}. In Eq. \ref{e:psop}, angular bracket denotes
that the average is taken over the $N_{cell}$ and the steady state configurations of the 
system. The $\Phi(f_a,\tau_p)$ estimates mixing or demixing of $A$ and $B$ particles in the
system: if a value of the $\Phi(f_a,\tau_p)$ is smaller than its critical value, then the 
system will be in the mixed state, and if $\Phi(f_a,\tau_p)$ is larger than its critical
value, then the system will be in the demixed state. Hence, we consider it as a phase
separation order parameter for this study.

\begin{figure}
	\includegraphics[width=8.0cm, height=7.0cm]{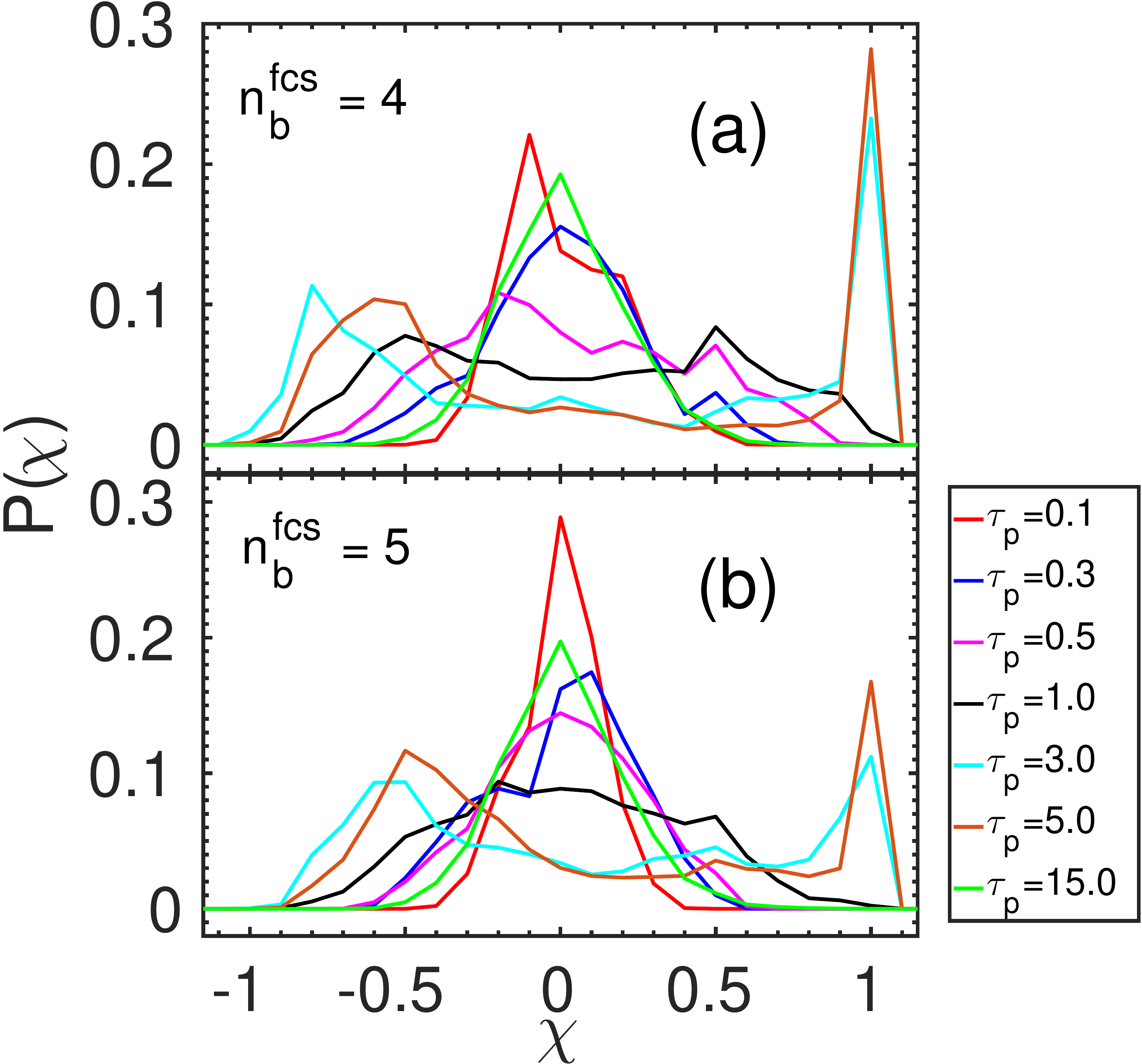}
	\caption{\label{f:pchi} Probability distribution of disparity in the number of 
	$A$ and $B$ particles, $P(\chi)$, at $f_a=$ 3.0, averaged over the square cells 
	and the steady-state configurations of the system. (a) $n_b^{fcs}=$ 4 and 
	(b) $n_b^{fcs}=$ 5.}
\end{figure}

Onset of the phase separation is considered at a critical value of the
$\Phi(f_a,\tau_p)$, denoted by $\Phi_c(f_a,\tau_p)$ for this study. To obtain 
the $\Phi_c(f_a,\tau_p)$ at activity $f_a$ and persistence time $\tau_p$, we compute
disparity in the number of $A$ and $B$ particles in each cell, which is calculated 
as $\chi=(n_A^i-n_B^i)/(n_A^i+n_B^i)$ over the steady states of the 
system \cite{j:activ_phassep_chandan}; the division of the simulation 
box into square cells is described above. Then, we compute distribution 
of $\chi$ as $P(\chi)$, which is shown in Fig. \ref{f:pchi}(a) and Fig. \ref{f:pchi}(b)
for $n_b^{fcs}=$ 4 and 5, respectively. Figure \ref{f:pchi}(a) depicts that $P(\chi)$ has a single
larger peak around $\chi=$ 0 at $\tau_p=$ 0.1, its height starts decreasing as the $\tau_p$ 
increases at $f_a=$ 3. The decrease in the peak height of the $P(\chi)$ accompanies with its
broadening that splits into two peaks at intermediate persistence times. The broadening in
$P(\chi)$ starts from $\tau_p=$ 0.3 and $f_a=$ 3 of $n_b^{fcs}=$ 4 [see Fig. \ref{f:pchi}(a)],
therefore, we consider the value of $\Phi(f_a,\tau_p)$ at $\tau_p=$ 0.3 as the onset value of
the phase separation order parameter, and that is $\Phi_c(f_a,\tau_p)=$ 0.3.
Bimodality due to the splitting of the main peak in $P(\chi)$ is
a clear sign of the disparity in the number of $A$ and $B$ particles in the square cells, 
and thus the signature of the phase separation. As the $\tau_p$ increases further, both peaks in 
Fig. \ref{f:pchi}(a) grow, and extent to the phase separation is shown (for example) 
at $\tau_p=$ 3 and 5 of $f_a=$ 3, where both peaks in $P(\chi)$ separate completely and their 
height also increases. From $\tau_p=$ 10 onwards, both peaks of $P(\chi)$ start merging again,
resulting in the single peak that exhibits the mixing of active and passive particles.
For $n_b^{fcs}=$ 5 [see Fig. \ref{f:pchi}(b)], the bimodality in $P(\chi)$ starts at the higher
$\tau_p$ than that for $n_b^{fcs}=$ 4, and the extent of phase separation is reduced because of
the smaller $\rho_{ab}$, which in turn reduces the peak height of $P(\chi)$ at $\tau_p=$ 3 and 5.
This system attains $\Phi_c(f_a,\tau_p)=$ 0.3, at the higher $\tau_p$ that is at $\tau_p=$ 0.5, 
where demixing starts in the system for the $n_b^{fcs}=$ 5. Thus, the extent to the phase 
separation is related to the number density of the active $B$ particles in this study.

\begin{figure}
	\includegraphics[width=8.0cm, height=7.0cm]{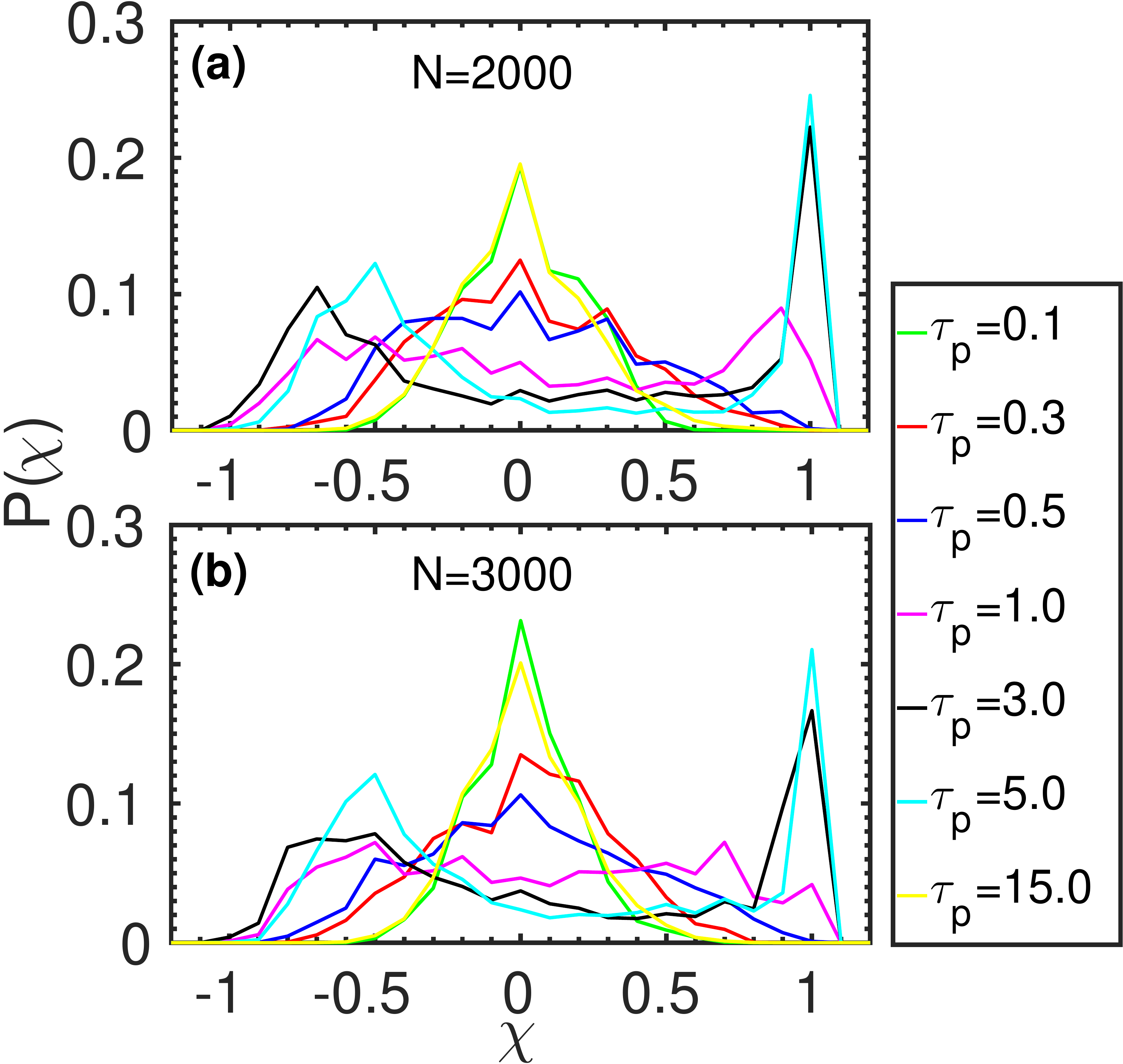}
	\caption{\label{f:pchise} Probability distribution, $P(\chi)$, at $f_a=$ 3.0 and local
	neighboring threshold $n_b^{fcs}=$ 4, averaged over the square cells 
	and the steady-state configurations of the system. (a) $N=$ 2000 and (b) $N=$ 3000
	particles.}
\end{figure}

To look at finite-size effects on the phase separation, we calculate the $P(\chi)$ for
$N=$ 2000 and 3000 particles; the procedure to compute $P(\chi)$ is the same as explained above.
Figure \ref{f:pchise} shows that the phase separation in the larger systems also starts at the 
same persistence time of the activity, which is examined from the broadening and splitting of
the main peak of the $P(\chi)$ as in the case of $N=$ 1000 particles system. 
Figure \ref{f:pchise} shows that the broadening in the $P(\chi)$ is more pronounced for the 
larger systems as compared to the $N=$ 1000 particles system, as size of domains due to the 
phase separation are different in the larger size systems compared to the smaller system at 
the same phase separation state points. 

\begin{figure}
	\includegraphics[width=8.5cm, height=7.0cm]{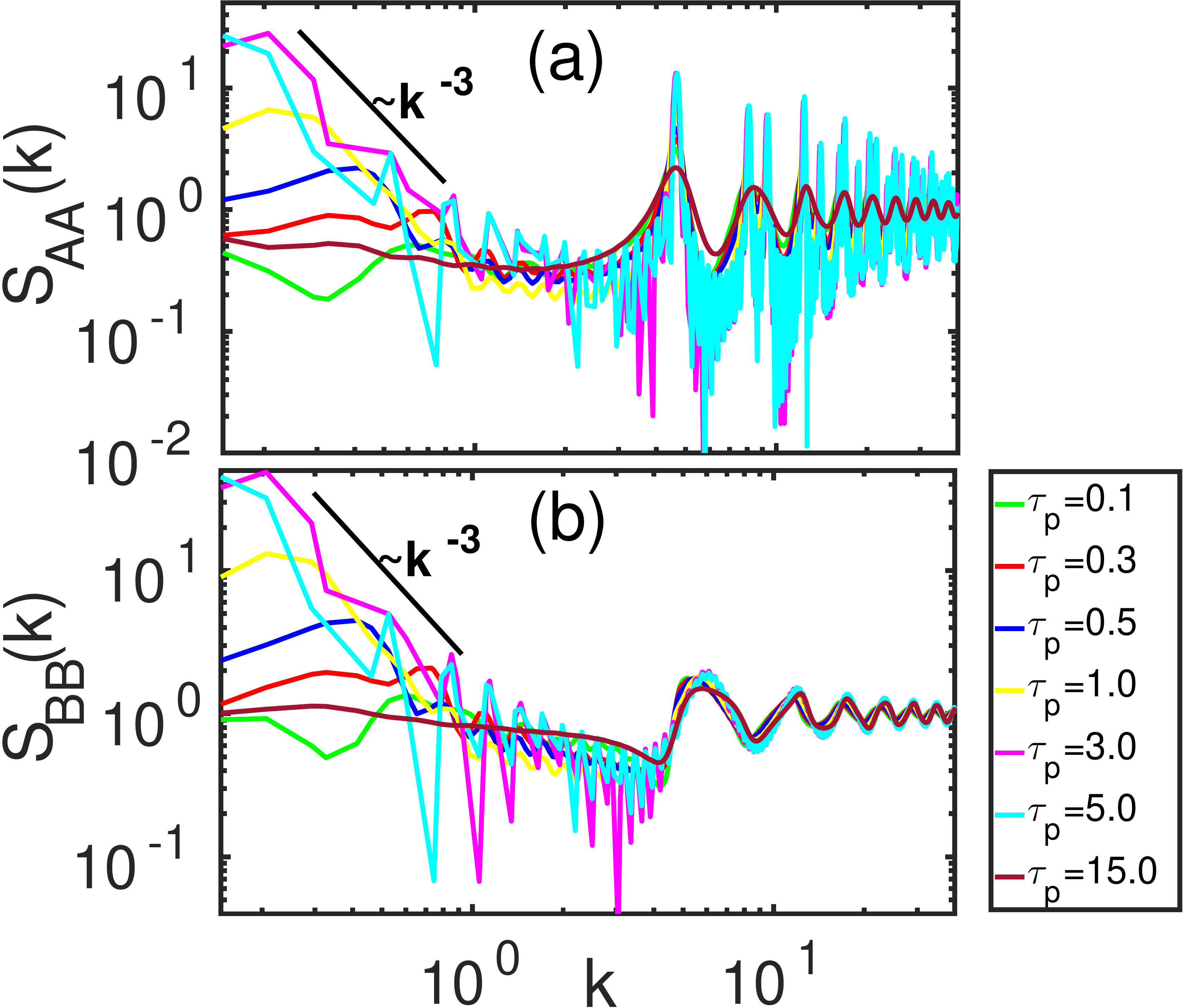}
	\caption{\label{f:sk} Static structure factor of (a) large and (b) small particles at the 
	activity $f_a=$ 3.0 and $n_b^{fcs}=$ 4. $S_{AA}(k)$ and $S_{BB}(k)$ is fitted with the Porod's
	law \cite{porod} of the form $S(k)\propto k^{-d-1}$ (see black lines) at $\tau_p=$ 1.0, 3.0 and
	5.0. Here, $d$ ($=$ 2) is the dimensionality of the system. The growth at low wave numbers
	in $S_{AA}(k)$ and $S_{BB}(k)$ is a signature of phase separation in the system \cite{furukawa}.}
\end{figure}

Further, the onset of the phase separation is also examined from the traditional correlation
function that is the static structure factor \cite{ham}, which is calculated as 
\begin{equation}
	\label{e:sk}
	S_{\alpha\beta}(k) = \frac{1}{N} \left\langle \sum_{i=1}^{N_\alpha}
	\sum_{j=1}^{N_\beta} \exp[-i\boldsymbol k.(\boldsymbol r_j - \boldsymbol r_i)]  \right\rangle,
\end{equation}
where $\boldsymbol k$ is a wave vector and $(\alpha,\beta)\in(A,B)$; the smallest wave number 
is given as $k=2\pi/L$. Here, we calculate structure factor of the $A$ and $B$ particles 
at $f_a=$ 3.0 and $n_b^{fcs}=$ 4 with varying persistence times, which is shown in Fig. \ref{f:sk}.
Structure factors of $A$ and $B$ particles are denoted by $S_{AA}(k)$ and $S_{BB}(k)$, respectively.
Figures \ref{f:sk}(a) and \ref{f:sk}(b) depict a growth in low wave number peak of $S_{AA}(k)$ 
and $S_{BB}(k)$, starting from the $\tau_p=$ 0.3. This growth in the peak of $S_{AA}(k)$ and 
$S_{BB}(k)$ at low $k$ is more pronounced at $\tau_p=$ 3.0 and 5.0, which corresponds to the extent
of the phase separation as examined from the splitting of the main peak of the $P(\chi)$ in two 
peaks [see Fig. \ref{f:pchi}(a)]. The growth in the peak of $S_{AA}(k)$ and $S_{BB}(k)$, at the
low wave numbers, is fitted with the Porod's law \cite{porod} of the form $S(k)\propto k^{-3}$ 
at $\tau_p=$ 1.0, 3.0, and 5.0, which is a signature of the phase separation in the 
system \cite{furukawa}. At $\tau_p=$ 15.0, the low $k$ peaks in $S_{AA}(k)$ and $S_{BB}(k)$ 
vanishes, which correspond to the continuous mixed state of the active-passive mixture, as
evidenced by the single peak in the $P(\chi)$ [see Fig. \ref{f:pchi}(a)]. Thus, our analysis 
of the phase separation from the $P(\chi)$ is corroborated by the growth of the low wave number
peaks in the structure factor of $A$ and $B$ particles.

\begin{figure}
	\includegraphics[width=8.5cm, height=7.5cm]{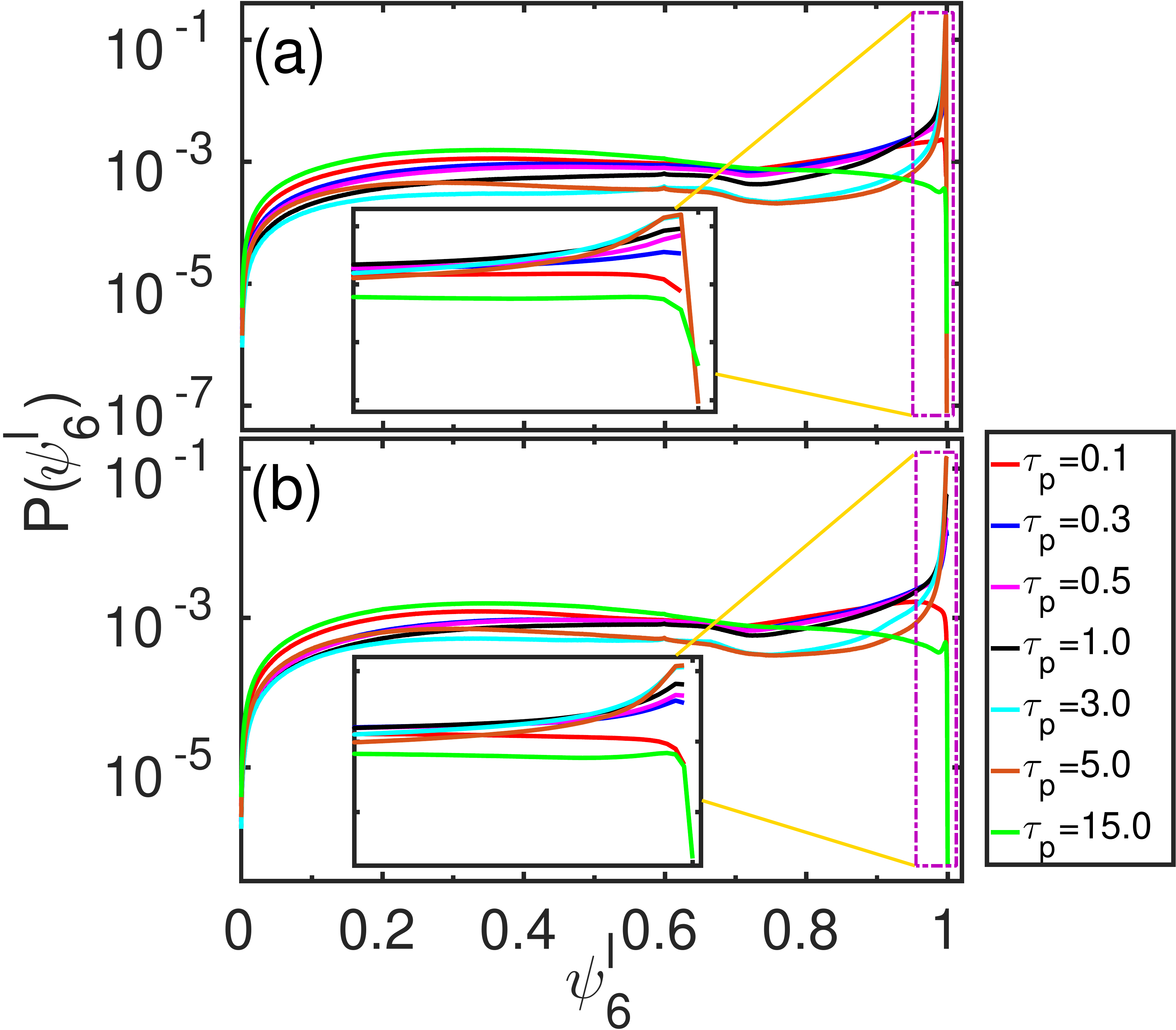}
	\caption{\label{f:psi6l} Probability distribution of local hexatic 
	order, $P(\psi_6^l)$, at $f_a=$ 3.0 and persistence times $\tau_p=$ 0.1-15.0. (a) 
	$n_b^{fcs}=$ 4 and (b) $n_b^{fcs}=$ 5. The insets in (a) and (b) show the
	zoomed (dash-dotted) area of $P(\psi_6^l)$ at $\psi_6^l=0.96-$ 1.0 . }
\end{figure}

As we know from the configurations (shown in Fig. \ref{f:hexpop}) that phase separation in the 
system coincides with the local hexatic ordering in $A$ particles. Therefore, we look
at a distribution of local hexatic ordering, averaged over the steady-state configurations
of the system. Distribution of single-particle hexatic order parameter for $A$ (large) 
particles, $P(\psi_6^l)$, is shown in Figs. \ref{f:psi6l}(a) and \ref{f:psi6l}(b) 
at $n_b^{fcs}=$ 4 and 5, respectively. Figure \ref{f:psi6l}(a) shows 
that $P(\psi_6^l)$ exhibits a peak 
near $\psi^l_6\simeq$ 1, which increases with the persistence time of the active force
up to $\tau_p=$ 5.0, at activity $f_a=$ 3.0. The peak in $P(\psi_6^l)$ is higher 
at $\tau_p=$ 3.0 and 5.0, where the active-passive mixture shows the extent of the phase
separation. From $\tau_p=$ 10 onwards, the peak of $P(\psi_6^l)$ near $\psi^l_6\simeq$ 1 
start decreasing, showing smaller values of $P(\psi_6^l)$ than that at $\tau_p=$ 0.1. 
This is because, at the higher $\tau_p$, the system prefers mixing and the hexatic order of 
majority of the particles decrease [see Fig. \ref{f:hexpop}(d)], which is also supported 
by the merging of double peaks of $P(\chi)$ into the single peak from $\tau_p=$ 10 
onwards [see Fig. \ref{f:pchi}(a) and Fig. \ref{f:pchi}(b)]. 
At $n_b^{fcs}=$ 5 [see Fig. \ref{f:psi6l}(b)], $P(\psi_6^l)$ is qualitatively similar 
to what found at $n_b^{fcs}=$ 4, only the extent to the orientational ordering 
in $A$ particles is smaller because the number density of the active $B$ particles reduces.
On comparing the $P(\psi_6^l)$ with the $P(\chi)$ at $f_a=$ 3, we show that the increment 
in the phase separation coincides with the increment in the hexatic ordering in $A$ 
particles. Thus, the extent to the phase separation in the system, the number density of 
active $B$ particles, and the hexatic order in the large particles are highly correlated.

\subsection{Phase diagram\label{s:pd}}

\begin{figure}
	\includegraphics[width=8.7cm, height=8.0cm]{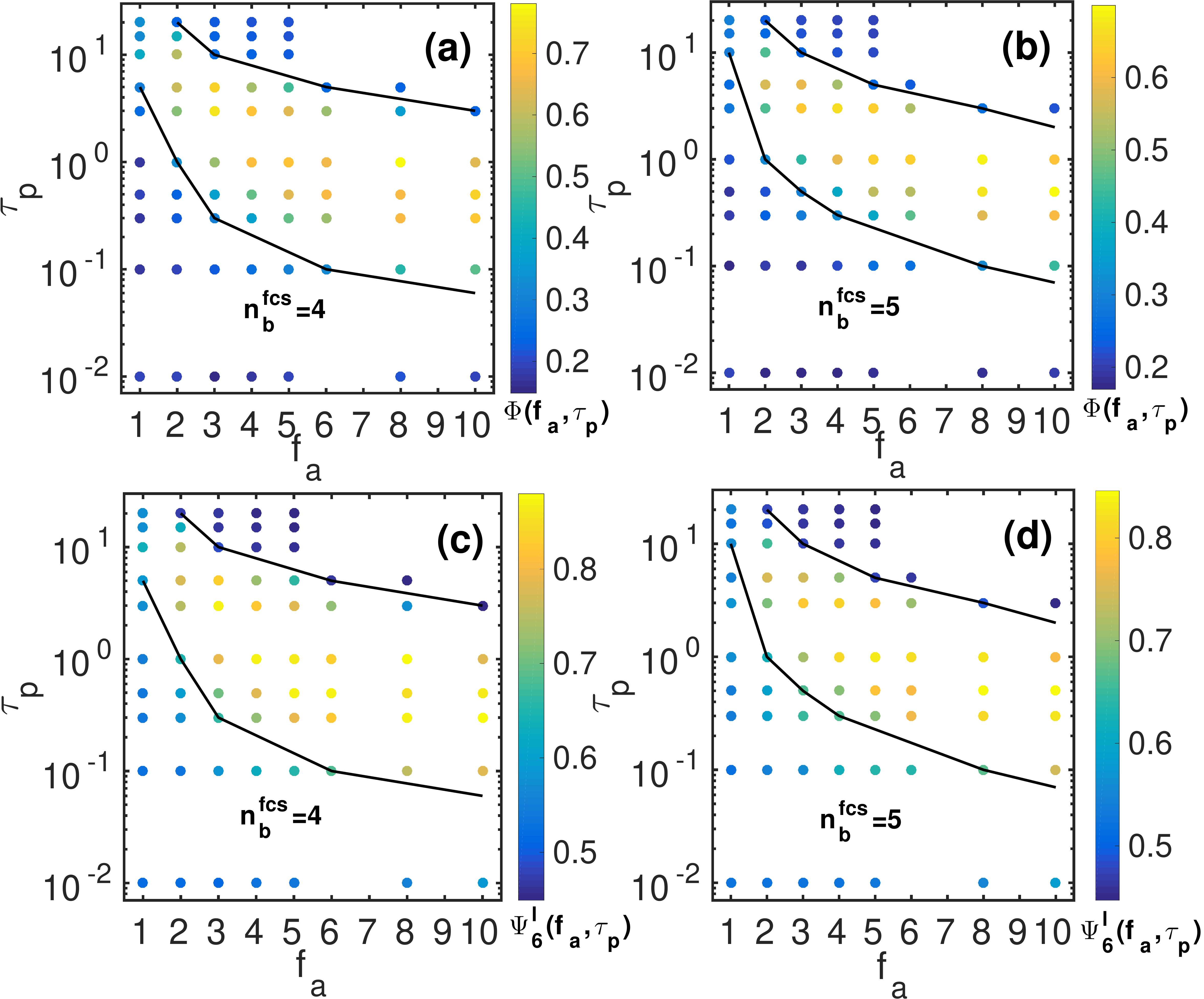}
	\caption{\label{f:phasdiag} Phase diagram of 2D active-passive binary mixture in 
	$\tau_p-f_a$ plane. (a--b) Order parameter of phase separation, $\Phi(f_a,\tau_p)$.
	(c--d) Average hexatic order parameter for large ($A$) particles, $\Psi_6^l(f_a,\tau_p)$.
	Black lines are corresponding to a critical value of $\Phi_c(f_a,\tau_p)=$ 0.3. }
\end{figure}

Figure \ref{f:phasdiag} shows a phase diagram of the active-passive binary mixture
in $\tau_p$-$f_a$ plane. To identify various regimes in the phase diagram, we calculate the 
phase separation order parameter $\Phi(f_a,\tau_p)$ [see Eq. \ref{e:psop} for its definition] 
and average hexatic order parameter for the large particles $\Psi_6^l(f_a,\tau_p)$.
An average absolute value of the hexatic order parameter is 
calculated as $\Psi_6 = N_\alpha^{-1} \sum_{j=1}^{N_\alpha} |\psi_6^j|$, where
$N_\alpha$ is the number of particles of type $\alpha \in (A,B)$ and $\psi_6^j$ is
defined in Eq. \ref{e:psi6}. Figures \ref{f:phasdiag}(a) and \ref{f:phasdiag}(b) show
that $\Phi(f_a,\tau_p)$ increases with $\tau_p$ up to its intermediate values at the
activity $f_a$. Further increasing the $\tau_p$, the $\Phi(f_a,\tau_p)$ start 
decreasing at each activity presented in the study. As described in Sec. \ref{s:pso}, 
the phase separation is highly correlated with the hexatic orientational ordering, 
therefore, we compute $\Psi_6^l(f_a,\tau_p)$ that is shown in
Figs. \ref{f:phasdiag}(c) and \ref{f:phasdiag}(d).
Similar to the $\Phi(f_a,\tau_p)$, the $\Psi_6^l(f_a,\tau_p)$ also increases with
the $\tau_p$ up to its intermediate values. Further increasing the $\tau_p$, 
$\Psi_6^l(f_a,\tau_p)$ decreases to even less than that its values at the smaller $\tau_p$.
This is because the active system is in its mixed state with smaller local hexatic order at
the higher $\tau_p$ [see Fig. \ref{f:hexpop}(d)]. Thus, the phase diagram 
at $n_b^{fcs}=$ 4, shown in Fig. \ref{f:phasdiag}(a) and Fig. \ref{f:phasdiag}(c), depicts
three regimes separated by the two transition lines: the bottom transition line (BTL) separates
the mixed state and the phase separation regime. The top transition line (TTL) of the phase 
diagram separates the phase separation region and the mixed state of active-passive
particles. Thus, it shows two transitions: one from mixing to demixing and another
from demixing to mixing again, which is the reentrant behavior of the system. Note that the 
mixed state of the particles above TTL is dissimilar to the mixed state of the particles
below the BTL because the former has less orientational ordering compared to the later.
The points on the BTL and TTL in the phase diagram [see Figs. \ref{f:phasdiag}(a--d)],
at each $f_a$ along the line of persistence time $\tau_p=$ 0.01-20, correspond to
the $\Phi_c(f_a,\tau_p)=$ 0.3. From the phase diagram, it is evident that the growth
of $\Phi$ is highly correlated with $\Psi_6^l$, up to the TTL from below. However, 
the $\Psi_6^l$ decreases above the TTL compared to below BTL because activity reduces the
local hexatic order that can be visualized by comparing Figs. \ref{f:hexpop}(a--d).
Now, it is tempting to identify the phases in these three regimes of the phase diagram by
computing positional and orientational correlations.

\subsection{Translational and orientational order}
Two-dimensional solids melt via a two-step process with the intermediate hexatic 
phase followed by the liquid phase. This melting occurs if volume 
fraction (or density) is decreased at a constant temperature, and agrees with the 
Kosterlitz, Thouless, Halperin, Nelson, and Young (KTHNY) two-step 
framework \cite{j:kosterlitz_thouless,j:halperin_nelson,j:young} in which solid to
hexatic and hexatic to liquid transitions are continuous.
According to the KTHNY theory, the solid phase is characterized by quasi-long-range 
positional order and proper long-range orientational order, whereas the hexatic phase is 
characterized by quasi-long-range orientational order and short-range positional
order; solids show the quasi-long-range positional order in two dimensions due to the 
Mermin-Wagner long-wavelength fluctuations \cite{j:mermin}. However, studies on melting
of the two-dimensional solids, consisting of repulsive hard disks \cite{j:2dmelt_hard_krauth} 
and steeply soft disks \cite{j:2dmelt_soft_krauth}, show solid to hexatic continuous
transition and hexatic to liquid first-order transition, within the KTHNY framework.
Therefore, in this study, we compute translational and orientational correlation functions
to identify the phases in the three regimes of the phase diagram. 

\begin{figure}
	\includegraphics[width=8.0cm, height=7.0cm]{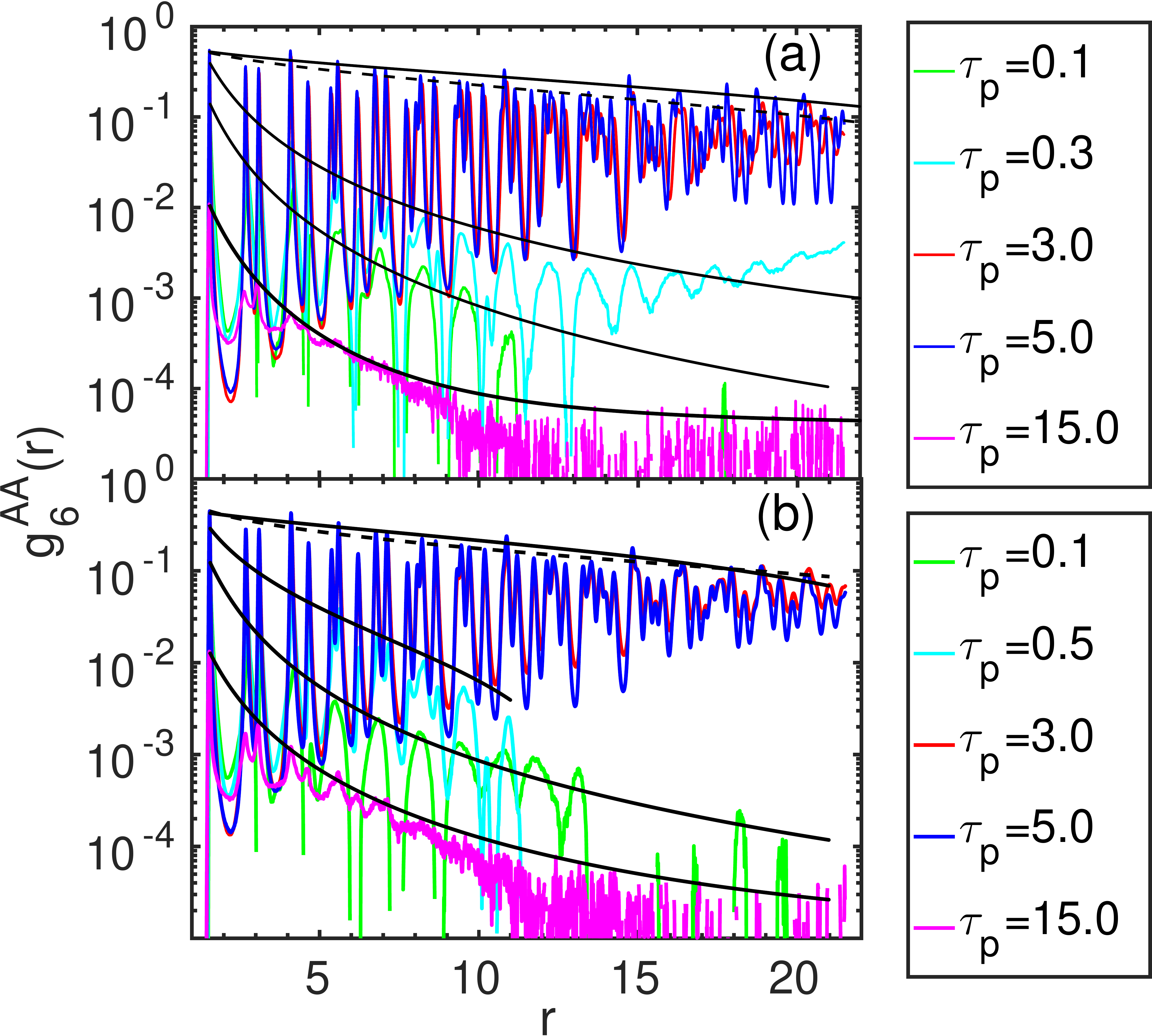}
	\caption{\label{f:g6rl} Hexatic order correlations in (a) at $n_b^{fcs}=$ 4
	and (b) at $n_b^{fcs}=$ 5. The fitting of $g_6^{AA}(r)$ in (a) 
	shows a power law $g_6^{AA}(r)\propto r^{-\eta6}$ with the exponents $\eta6=$ 2.78, 2.26, 
	0.035, 0.014, 2.89 at $\tau_p=$ 0.1, 0.3, 3.0, 5.0, 15.0, respectively. The fitting of
	$g_6^{AA}(r)$ in (b) shows the power law with exponents $\eta6=$ 2.68, 1.53, 
	0.15, 0.19, 2.53 at $\tau_p=$ 0.1, 0.5, 3.0, 5.0, 15.0, respectively.}
\end{figure}

In Sec. \ref{s:pso} and Sec. \ref{s:pd}, we have computed $P(\psi_6^l)$ and $\Psi_6^l$
to examine the average local hexatic order distribution and the average hexatic order 
parameter for the large particles, respectively. To look at the range of this ordering, 
we compute hexatic order correlations as
\begin{equation}
	\label{e:g6r}
	g_6(r)= \frac{\langle \psi_6^j {\psi_6^k}^*
	\rangle|_{\mathbf r_j-\mathbf r_k=r}}{\langle|\psi_6^j|^2\rangle}.
\end{equation}
A plot of $g_6^{AA}(r)$ and its fitting are displayed 
in Fig. \ref{f:g6rl}(a) and Fig. \ref{f:g6rl}(b) corresponding to $n_b^{fcs}=$ 4 and
$n_b^{fcs}=$ 5, respectively. Figure \ref{f:g6rl}(a) shows that the $g_6^{AA}(r)$ is fitted 
with the power law, $g_6^{AA}(r)\propto r^{-\eta6}$, with an exponent $\eta6$ that decreases
continuously on increasing the $\tau_p$ up to $\tau_p=$ 5.0 at $f_a=$ 3.0. 
At the $\tau_p=$ 3.0 and 5.0, the exponent $\eta6$ reaches, respectively at the values of 
0.03 and 0.01 that are far smaller than 1/4, the stability limit of the hexatic phase
in the continuous KTHNY transition from solid to hexatic phase. According to the KTHNY theory, 
there should be no decay of $g_6(r)$ in the solid phase that corresponds to the proper long-range
orientational order. Thus, due to very small fitting exponents of the $g_6^{AA}(r)$ at the 
$\tau_p=$ 3.0 and 5.0, we call it a solidlike phase. On further increment in the $\tau_p$ at 
the activity $f_a=$ 3.0 (above the TTL) that is $\tau_p=$ 10.0 onwards, the $g_6^{AA}(r)$ decays 
rapidly [see Fig. \ref{f:g6rl}(a)], and the exponent $\eta6$ again increases to 2.89, though 
it does not fit using simple exponential that shows short-range-like orientational order 
in $A$ particles, corresponds to the active liquid phase; the corresponding configuration is
shown in Fig. \ref{f:hexpop}(d). This is because the solidlike phase melts to the hexatic that
further melts to the liquid, and the orientational ordering reduces.
For $n_b^{fcs}=$ 5 [see Fig. \ref{f:g6rl}(b)], the qualitative nature of the $g_6^{AA}(r)$ is similar,
though it differs quantitatively because the number density of the active $B$ particles is smaller 
that causes less phase separation. The exponents of the fitting of $g_6^{AA}(r)$ decrease as
$\eta6=$ 0.15 and 0.19 at the $\tau_p=$ 3.0 and 5.0, however, they are within the range
of $0<\eta6<1/4$, which manifests that they correspond to the (quasi-)long-range orientational
order in the $A$ particles for $n_b^{fcs}=$ 5.

\begin{figure}
	\includegraphics[width=6.5cm, height=6.0cm]{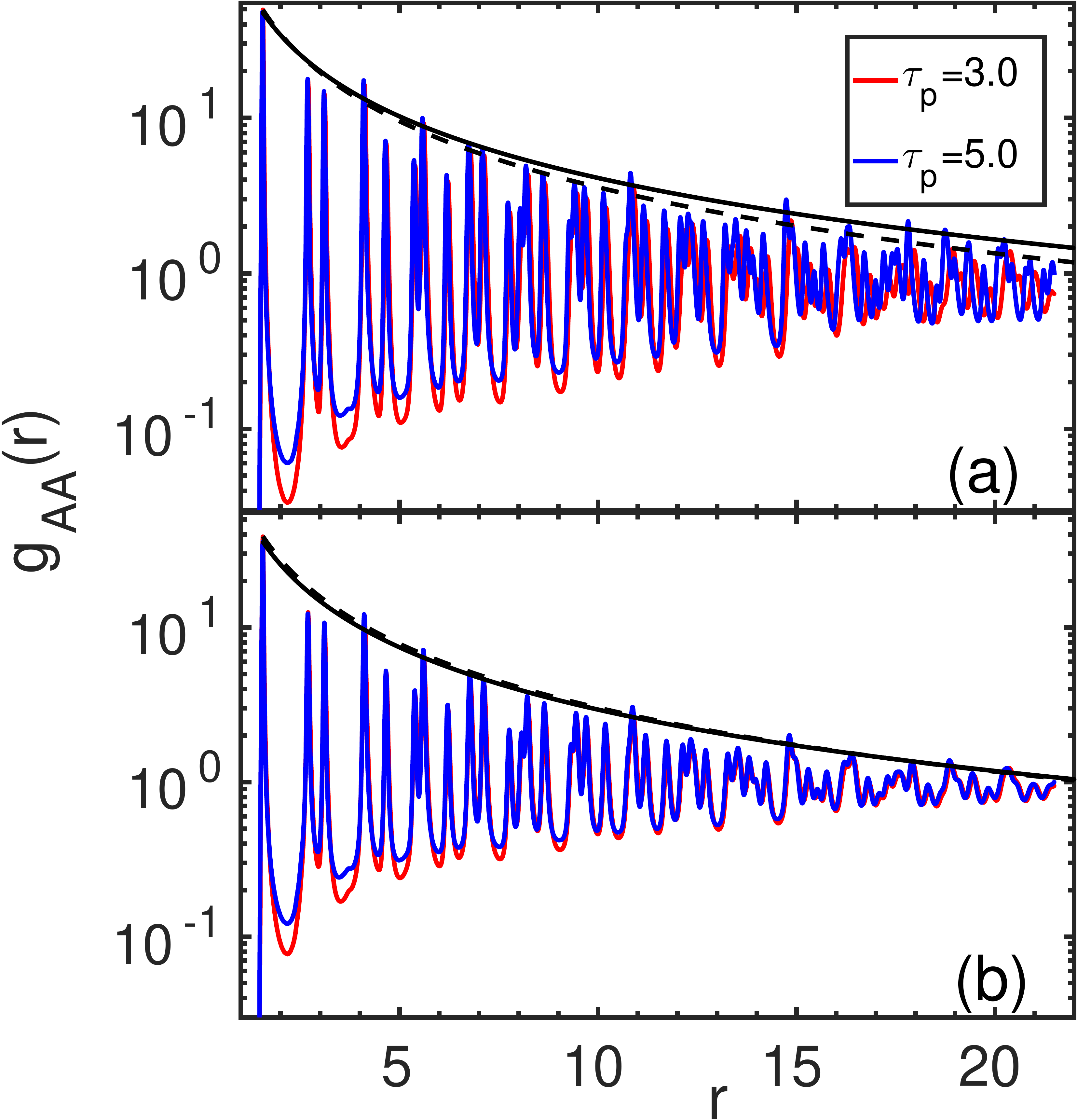}
	\caption{\label{f:gr} Radial distribution function of large particles in
	(a) $n_b^{fcs}=$ 4 and (b) $n_b^{fcs}=$ 5. The $g_{AA}(r)$ is fitted with the power 
	law of the form, $g_{AA}(r)\propto r^{-\eta}$, which shows (quasi-)long-range positional 
	order. }
\end{figure}

\begin{figure}
	\includegraphics[width=8.0cm, height=6.5cm]{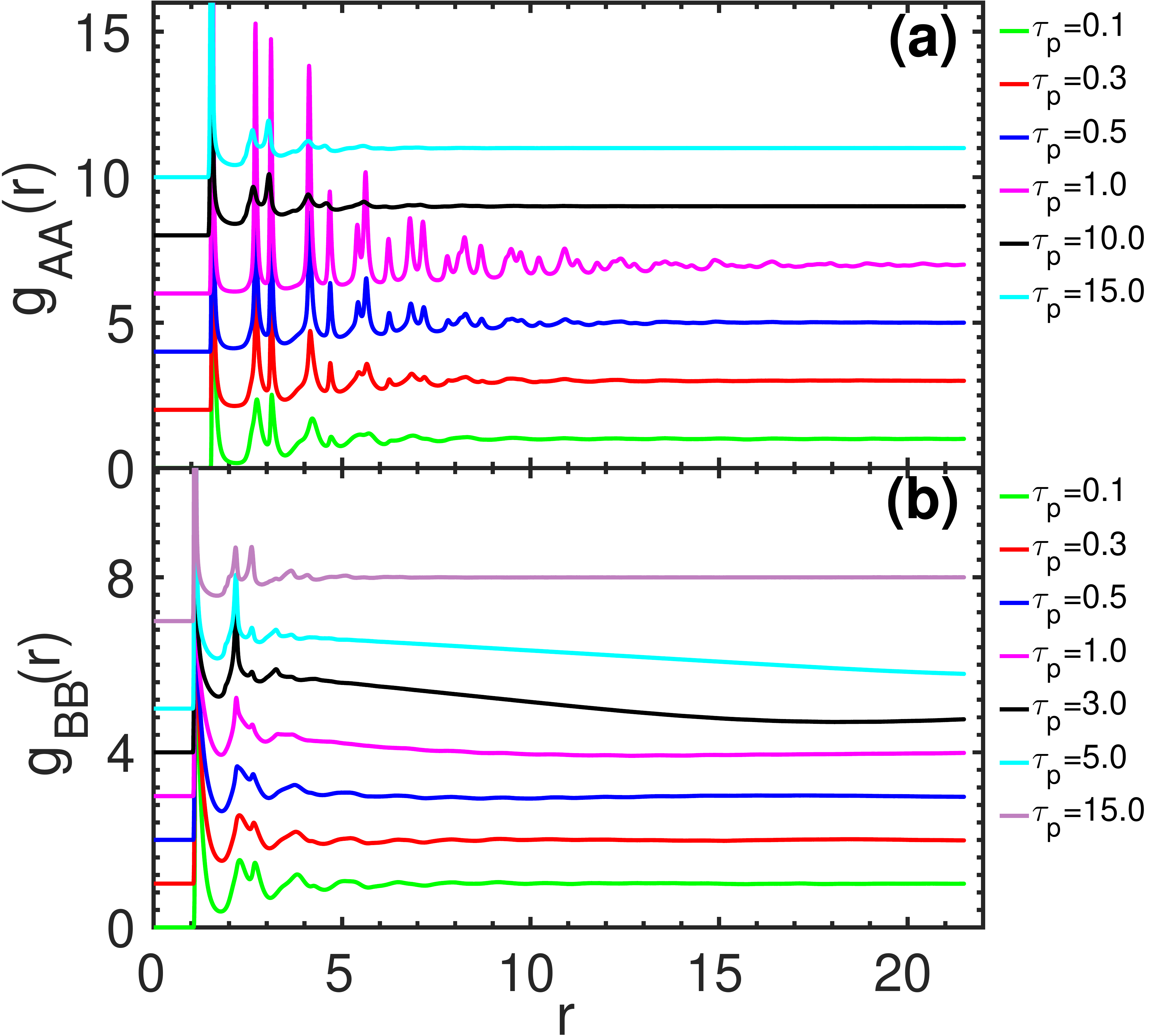}
	\caption{\label{f:gract} RDF of (a) $AA$ and (b) $BB$  
	particles at $f_a=$ 3.0 for $n_b^{fcs}=$ 4. In (a) and (b) plots are shifted
	by 2.0 and 1.0 along the $y-$axis for visibility, respectively.}
\end{figure}

To look at the positional order, we compute the RDF of $A$ and $B$ particles,
$g_{\alpha\beta}(r)$, defined as
\begin{equation}
	\label{e:gr}
	g_{\alpha\beta}(r) = \frac{A}{2\pi r \Delta r N_\alpha N_\beta} \left\langle
	\sum_{i=1}^{N_\alpha}\sum_{\substack{j=1 \\j \ne i}}^{N_\beta} 
	\delta({r - |\mathbf r_j - \mathbf r_i|}) \right\rangle,
\end{equation}
where $(\alpha,\beta) \in (A,B)$.
The average hexatic order [see Figs. \ref{f:phasdiag}(c) and \ref{f:phasdiag}(d)], the
distribution of large particles' local hexatic order [see Fig. \ref{f:psi6l}], and
the configurations of the system [see Fig. \ref{f:hexpop}], show that ordering in $A$ 
particles increases with the persistence time up to its intermediate values and start
decreasing at the higher $\tau_p$.
Therefore, we calculate RDF of $AA$ particles and look at its fitting to examine the
positional order comprising $A$ particles in the system.
Figures \ref{f:gr}(a) and \ref{f:gr}(b) show $g_{AA}(r)$ and its fitting 
for $n_b^{fcs}=$ 4 and 5, respectively. The $g_{AA}(r)$ is fitted with the power law of the 
form $g_{AA}(r)\propto r^{-\eta}$ at the $\tau_p=$ 3.0 and 5.0 that are corresponding to the
extent of the phase separation in the system at $f_a=$ 3.0. The obtained exponents 
are $\eta=$ 1.42 and 1.31 that are far larger than 1/3, the stability limit of the solid 
phase in the KTHNY continuous transition. However, the positional correlations decay 
exponentially in the hexatic phase, according to the KTHNY theory. Therefore, we consider 
the power-law decay of the $g_{AA}(r)$ like the quasi-long-range positional order 
consisting of $A$ (passive) particles. Thus, the passive $A$ particles are the solidlike 
at $\tau_p=$ 3.0 and 5.0 of $f_a=$ 3.0. The extent of the phase separation in the system
coincides with the phase separation of the solidlike phase consisting of only passive $A$
particles, as shown in Fig. \ref{f:hexpop}(c) for visualization.
Figure \ref{f:gract}(a) shows the RDF of $AA$ particles at the activity $f_a=$ 3.0
to look at the variation of positional order with persistence time in the range 
of $\tau_p=$ 0.1-15.0. The micro-structure of the passive (large) particles becomes rich with
the increase in $\tau_p$ till the formation of the solidlike phase (see RDF of Fig. \ref{f:gr}).
This is because the peaks in $g_{AA}(r)$ split with increment in their height and become
sharper. Thus, the RDF of $AA$ particles show that the positional order comprising $A$ particles 
grows from the short-range order (at smaller $\tau_p$) to the quasi-long-range order like at 
the intermediate $\tau_p$ of the active force. Further increase in the $\tau_p$, reduces the 
positional order in the $A$ particles, which is due to the melting of the solidlike phase into
the hexatic phase that further melts to the liquid phase.

Now, we look at the structural changes in the $B$ particles
(active and passive) of the binary mixture with the application of activity.
The RDF of $BB$ particles [see Fig. \ref{f:gract}(b)] shows the reduction in its
micro-structure as the $\tau_p$ is increased from the passive limit ($f_a=$ 0) of 
the activity, till the TTL is reached. The micro-structures comprising $B$ particles
are considered as the splitting of the secondary peak
of $g_{BB}(r)$ \cite{j:kobsplt}, which start merging with increment in the
$\tau_p$ [see Fig. \ref{f:gract}(b)]. Interestingly, the merging of secondary peaks'
split in $g_{BB}(r)$ coincides with the increase in the positional ordering in
$g_{AA}(r)$. At $\tau_p=$ 3.0 and 5.0 of the activity $f_a=$ 3.0, where the passive
particles comprise the solidlike phase, height of the peaks in 
the $g_{BB}(r)$ [see Fig. \ref{f:gract}(b)] decreases with $r$ up to $r=$ 4.0 that
oscillates around $g_{BB}(r)=$ 2.0. The $g_{BB}(r)$ goes below 1.0 at longer distances, 
showing the phase separation of $A$ and $B$ particles, where $B$ particles are phase
separated from the solidlike phase consisting of only $A$ particles. This coexistence
region is consisting of small ($B$) and few large particles. Above TTL, the secondary peak 
split in $g_{BB}(r)$ reappears, shows mixing of the particles in the active liquid phase. 
Thus, the RDF of $A$ and $B$ particles show that the positional ordering increases in $A$ 
particles, whereas it decreases in $B$ particles, up to the intermediate $\tau_p$ (below TTL).
Above TTL, both $A$- and $B$-type particles show only short-range positional ordering, and
the system is in the liquid phase.

\begin{figure}
	\includegraphics[width=7.0cm, height=6.5cm]{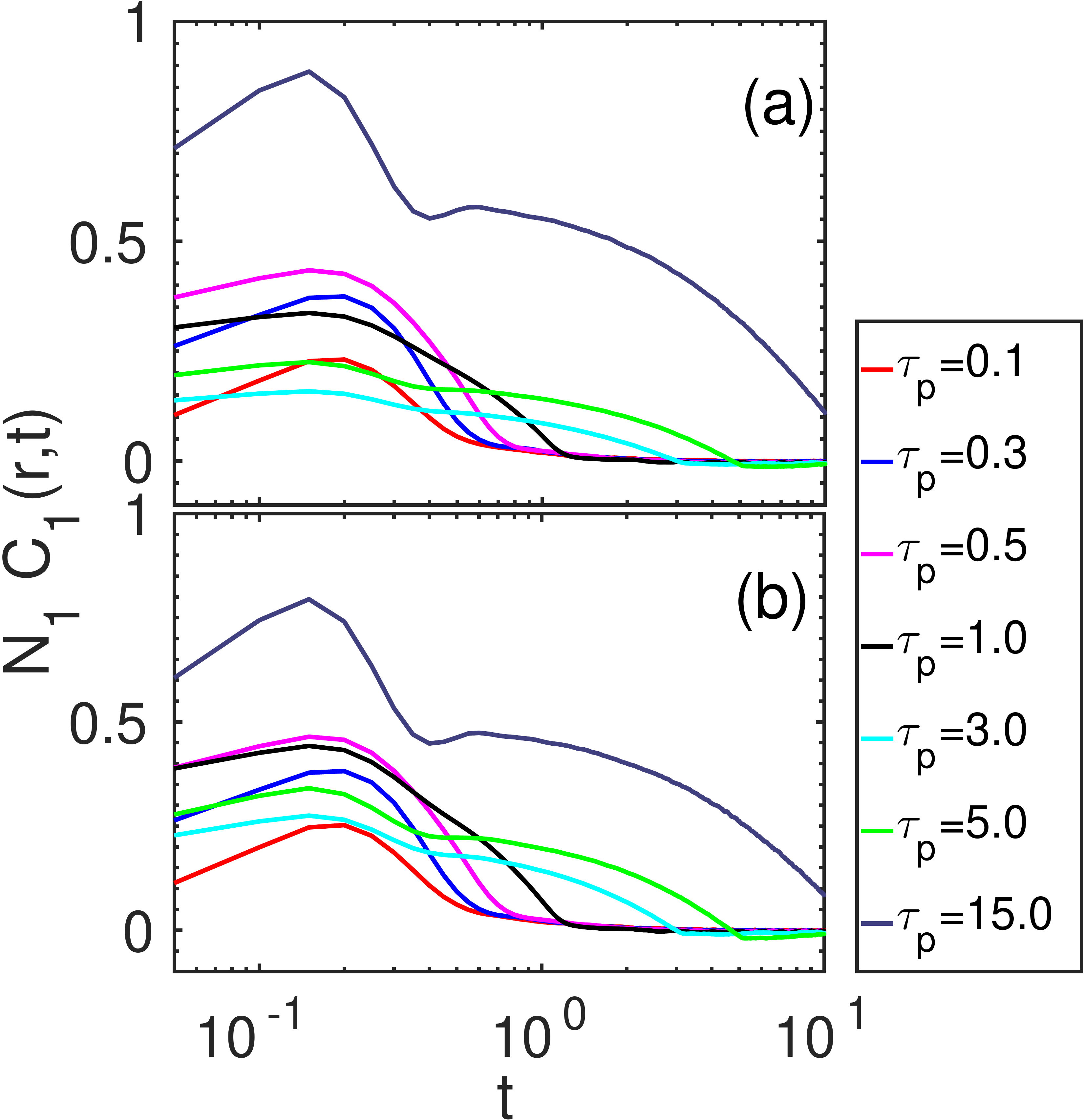}
	\caption{\label{f:vcc} Velocity cross-correlations ($B-A$) within the first
	coordination shell at the persistence	times $\tau_p=$ 0.1-15.0 at $f_a=$ 3.0.
	(a) $n_b^{fcs}=$ 4 and (b) $n_b^{fcs}=$ 5.}
\end{figure}

Thus, we have identified three regimes, namely, active glass (below BTL), active 
liquid (above TTL), and the phase separation (between the BTL and TTL) regimes by two order 
parameters: one is the phase separation order parameter $\Phi(f_a,\tau_p)$ and another is
the average hexatic orientational order parameter of the large 
particles, $\Psi_6^l(f_a,\tau_p)$. The solidlike, hexatic, and liquid phases are identified 
by the radial distribution functions and hexatic order correlations function. However, the 
active glass is identified by looking at the relaxation dynamics of the system, presented in
the Appendix \ref{s:glas_act}.

\subsection{Velocity cross correlations\label{s:vcc}}
The QS active particles inject energy into the system that induces the transition from 
the active glass (mixed state) to the solid-liquid phase coexistence to the active
liquid (mixed state) phases. To look at the reentrant in the orientational ordering of 
the system, with the persistence time of the active force, we calculate velocity 
cross correlations \cite{j:momtr_balucani,j:momtr_binmix} between $B$ and $A$ particles to
examine the momentum transfer between them, which is defined as
\begin{equation}
	C_n(r,t) = \frac{1}{N_n} \frac{{\left\langle \mathbf v_i(0).\mathbf 
	v_j(t)\right\rangle}_n}{\left\langle v_i^2(0) \right\rangle},
\end{equation}
where $N_n$ are the number of particles in the $n^{th}$ coordination shell.
Normalized momentum transfer within the FCS, $N_1C_1(r,t)$, is displayed in 
Fig. \ref{f:vcc}(a) and Fig. \ref{f:vcc}(b) at  $n_b^{fcs}=$ 4 and 5, respectively. 
Figure \ref{f:vcc}(a) shows that the momentum transfer between $B$ and $A$ particles is 
small at smaller $\tau_p$, which increases with it up to 
$\tau_p=$ 0.5. This is because the activity is insufficient to break the cages of the 
particles at the smaller $\tau_p$. The transfer of momentum from $B$ to $A$ particles 
decreases at $\tau_p=$ 1.0 and 3.0, it start increasing again from $\tau_p=$ 5.0, though 
it is less than $\tau_p=$ 0.3. At this intermediate $\tau_p$, activity fluidizes the 
glassy $B$ particles, whereas the $A$ particles that are passive start phase separating, 
resulting in the (quasi-)long-range positional and long-range orientational-like ordering.
At the intermediate $\tau_p$, activity does not destroy the ordering of the $A$ particles, 
instead enhance it, and the system phase separates. Figure \ref{f:vcc}(a) shows that the momentum
transfer between $B$ and $A$ particles is pronounced (from and above TTL) as the peak heights
of $N_1C_1(r,t)$ shoots up because the $\tau_p$ is large enough to destroy the ordering of the
$A$ (passive) particles. Thus, the resulting phase appears as the active liquid with the mixing
of $A$ and $B$ particles. At $n_b^{fcs}=$ 5, the $N_1C_1(r,t)$ is qualitatively similar,
though the momentum transfer between $A$ and $B$ particles is smaller than the $n_b^{fcs}=$ 4.

\section{Conclusion}

We have developed a model system for the dense active systems, where the activity is 
introduced by the local-density-dependent quorum sensing scheme. By applying activity to the small 
particles in the binary colloidal mixture, the system undergoes phase separation, accompanied by 
the ordering, at the intermediate persistence time of the activity. Thus, the active-passive binary
mixture shows a continuous growth of hexatic order consisting of passive (large) particles from
its passive glassy state on increasing persistence time at a constant activity, till the formation
of the solidlike phase. Further increasing the persistence time of the active force, the positional
order vanishes, and the hexatic order reappears that reduces with the persistence time. Finally, 
we have found that the stability in the solidlike phase is due to the least momentum transfer 
between active and passive particles in the phase separation regime.

\begin{acknowledgments}
	The authors acknowledge financial support from the Department of Atomic Energy, India 
	through the 12th plan project (12-R\&D-NIS-5.02-0100).  
\end{acknowledgments}

\appendix

\renewcommand{\thefigure}{A.\arabic{figure}}
\setcounter{figure}{0}

\section{Glassy features of passive system\label{s:glas_pasiv}}

\begin{figure}
	\includegraphics[width=8.0cm, height=3.5cm]{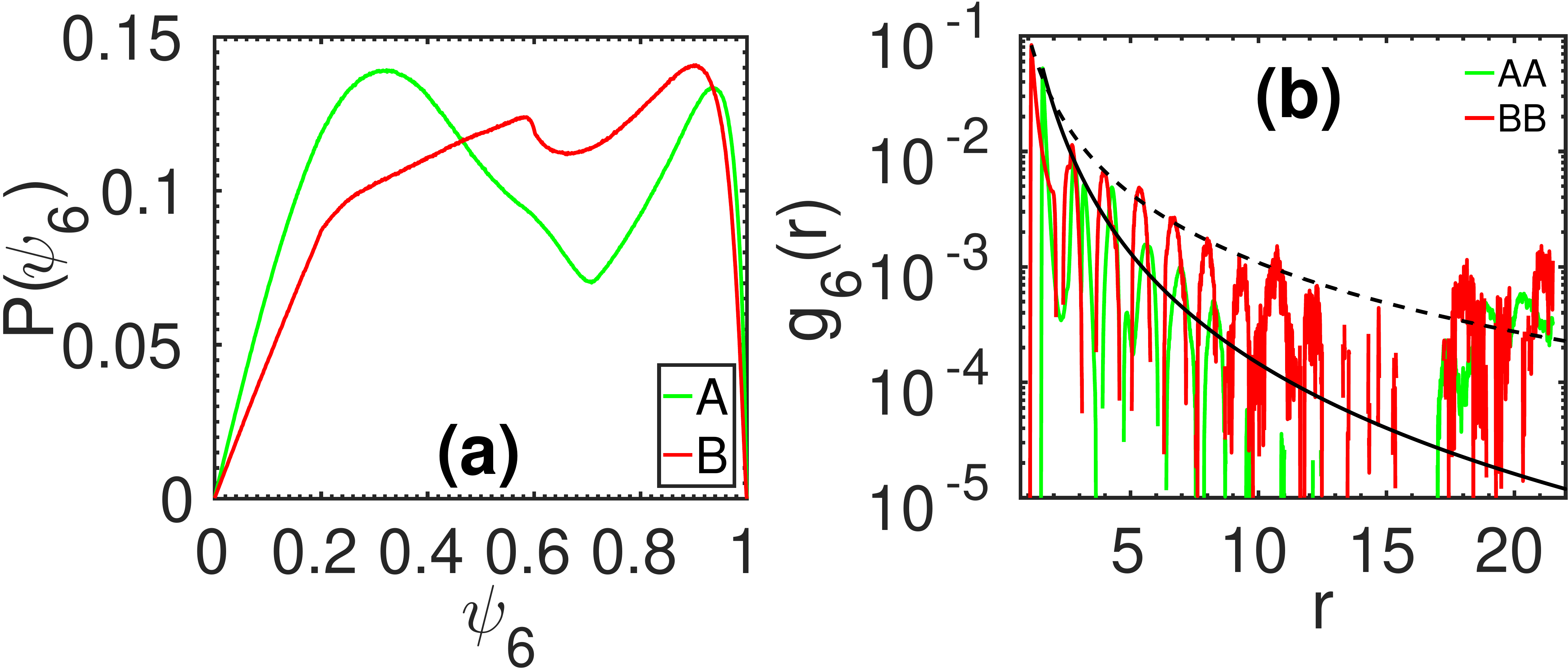}
	\caption{\label{f:hop} (a) Probability distribution $P(\psi_6)$ of $A$
	and $B$ particles averaged over the equilibrium configurations of the passive
	system at $T=$ 0.01, and (b) hexatic order correlation function $g_6(r)$ fitted with a
	power law of the form, $g_6(r)\propto r^{-\eta6}$; the exponent $\eta6=$ 3.2 and 2.0 
	for $AA$ (solid curve) and $BB$ (dashed curve) particles.}
\end{figure}

The average hexatic ordering in the passive colloidal mixture is examined by 
calculating probability distribution $P(\psi_6)$, averaged over the equilibrium
configurations of the system at $T=$ 0.01. The $P(\psi_6)$ of $B$
particles [see Fig. \ref{f:hop}(a)] shows a pronounced peak around $\psi_6=$ 0.9,
which exhibits an emergence of the hexatic order. However, the $P(\psi_6)$ of $A$ 
particles shows two peaks, one around $\psi_6=$ 0.32 and another around $\psi_6=$ 0.93,
which manifest the smaller orientational order in $A$ particles compared to the $B$ 
particles because $B$ particles show a smaller hump at the low values of the $\psi_6$.
To examine the range of hexatic orientational ordering, we compute the hexatic order 
correlation function $g_6(r)$, defined in Eq. \ref{e:g6r}, which is shown in 
Fig. \ref{f:hop}(b). The $g_6(r)$ is fitted with the power law of the form
$g_6(r)\propto r^{-\eta6}$ with the exponents $\eta6=$ 3.2 and 2.0, respectively for $AA$ 
and $BB$ particles. The exponents are far larger than $1/4$, the stability limit of the 
hexatic phase in the KTHNY scenario. Thus, it shows that our passive binary mixture
consists of medium-range hexatic order, a feature of supercooled
liquids \cite{j:tanaka_2dglasorder,j:smrajit_mrco}.
The smaller value of the exponent $\eta6$ for $B$ particles compared to the $A$ 
particles means a larger correlation length of hexatic order for $B$ particles. Thus,
the passive system shows more (local) hexatic orientational ordering for the $B$ 
particles compared to the $A$ particles in this system.

\begin{figure*}
	\centering
	\includegraphics[width=14.0cm, height=3.8cm]{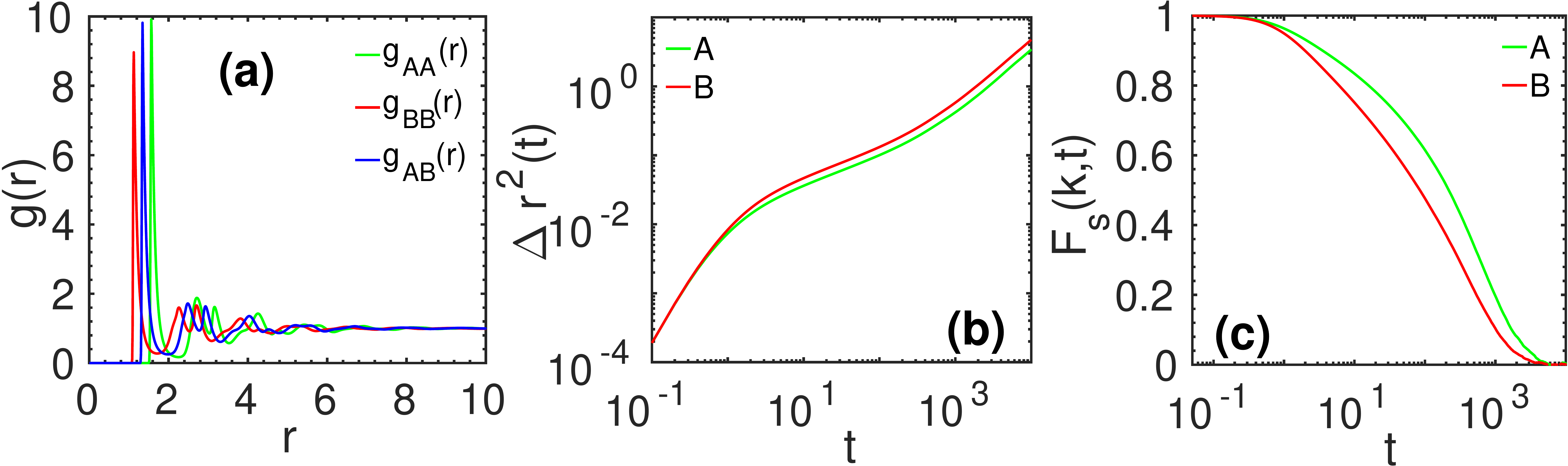}
	\caption{\label{f:stcdyn} Passive mixture at $T=$ 0.01 and the area fraction
	$\phi=$ 0.628: (a) RDFs of the binary colloidal mixture, (b) MSD of $A$ and
	$B$ particles, (c) incoherent intermediate scattering function, $F_s(k,t)$, at wave
	numbers corresponding to the first peak of static structure factor $S(k)$ of $A$ and
	$B$ particles, $k_A=$ 4.5 and $k_B=$ 5.0, respectively.}
\end{figure*}

The hexatic orientational order found in the absence of positional order at the 
size ratio 1:1.4 \cite{j:onuki_2dglass}, is termed as the glassy structural order
by Kawasaki and Tanaka \cite{j:tanaka_2dglasorder}. Therefore, we 
examine the positional order in the system, which can be obtained from the RDF given 
in Eq. \ref{e:gr} (or its Fourier transform, widely known as structure factor).
Figure \ref{f:stcdyn}(a) depicts the $g(r)$ of $AA$, $BB$, and $AB$ particles, which shows
an absence of long-range (or quasi-long-range) positional order in the system, however there
is a medium-range hexatic orientational order as shown in
Figs. \ref{f:hop}(a) and \ref{f:hop}(b). This confirms that the passive binary colloidal
mixture is in its amorphous state at $T=$ 0.01 and area fraction $\phi=$ 0.628.

The structure of the 2D binary mixture shows that it is in the amorphous (glassy) 
state as described in the previous section. To confirm this, we look at the 
dynamics of the system. The glassy dynamics of the passive binary colloidal mixture
is examined using the mean-squared displacements (MSD) and the incoherent intermediate
scattering function (IISF), which are defined as
\begin{equation}
	\Delta r^2(t) = \frac{1}{N}\left\langle \sum_{i=1}^N 
	[\mathbf r_i(t)-\mathbf r_i(0)]^2 \right\rangle
\end{equation} and 
\begin{equation}
	\label{e:fskt}
	F_s(k,t) = \frac{1}{N}\left\langle \sum_{j=1}^N 
	\exp\{i \mathbf k.[\mathbf r_j(t)-\mathbf r_j(0)]\} \right\rangle,
\end{equation}
where wave number $k$ corresponds to the first peak of static structure factor of $A$ and 
$B$ particles. Here, we use $k=$ 4.5 and 5.0 for $A$ and $B$ particles, respectively.
Figure \ref{f:stcdyn}(b) shows the MSD of $A$ and $B$ particles at $T=$ 0.01, which exhibits
a ballistic regime at short times that crossing over to the sub-diffusive regime around
$t\approx$ 1.0, becomes diffusive at the longer times. The different regimes in the MSD 
of $A$ and $B$ particles are identified from the slope 
$\alpha = \partial ln \Delta r^2(t) / \partial ln t$ :
at short times (ballistic regime) $\alpha \approx$ 2,
in sub-diffusive regime $0<\alpha<1$, while in the diffusive regime $\alpha=$ 1. The 
sub-diffusive regime appears due to the structural cages around a tagged 
particle by its neighboring particles \cite{j:berthier}. These structural cages cause
a slow down in the dynamics, thus the MSD decreases at this time scale. At long times,
the particles come out of these structural cages, thus MSD again shoots up and the dynamics 
becomes diffusive. Although the larger ($A$) particles are slower than the smaller ($B$) 
ones, both sizes of particles show the sub-diffusive regime at the intermediate times [see 
Fig. \ref{f:stcdyn}(b)], which is one of the hallmarks of the glassy 
dynamics \cite{j:berthier}. The structural relaxation of these cages is characterized 
using the IISF, $F_s(k,t)$ of both sizes of particles, which examines the relaxation of 
density fluctuations due to the structural cages of the particles at a wave number
$k$. In this passive binary colloidal mixture, $F_s(k,t)$ is fitted using an empirical
Kohlrausch-Williams-Watts (KWW) function of the form
$f(t)^{KWW} \propto \exp(-(t/\tau)^\beta)$, where $\beta$ is an
exponent \cite{j:berthier}. The value of the exponent $\beta=$ 1 for the exponential
density relaxations (liquid state), whereas $0<\beta<1$ for the non-exponential 
relaxations found in slow relaxing systems, e.g., glass-forming liquids, 
intracellular dynamics, motion in the crowded media, etc. For this passive 
binary mixture, we obtain $\beta=$ 0.66 and 0.6 for $A$ and $B$ particles, respectively.
These values of $\beta$ show the non-exponential relaxation dynamics of the system at
$T=$ 0.01 and area fraction $\phi=$ 0.628, which is a signature of the glassy dynamics
in the passive system.

\section{Relaxation dynamics of the active system\label{s:glas_act}}

Many characteristics of glassy systems are observed in the system of passive binary
disks using structure and dynamics; the structure of the active system changes with
the activity. As discussed in the phase diagram of the main paper, the system shows 
three regimes separated by the two transition lines: mixed region, phase separation,
and active liquid with the mixing of both types of particles. The phase separation and the
phases therein, and the active liquid are characterized in the main paper using various
observables. Here, we examine the dynamics of the active-passive mixture from smaller to
longer persistence times across both transition lines of the phase diagram at 
activity $f_a=$ 3.0.

\begin{figure}
	\includegraphics[width=7.5cm, height=7.5cm]{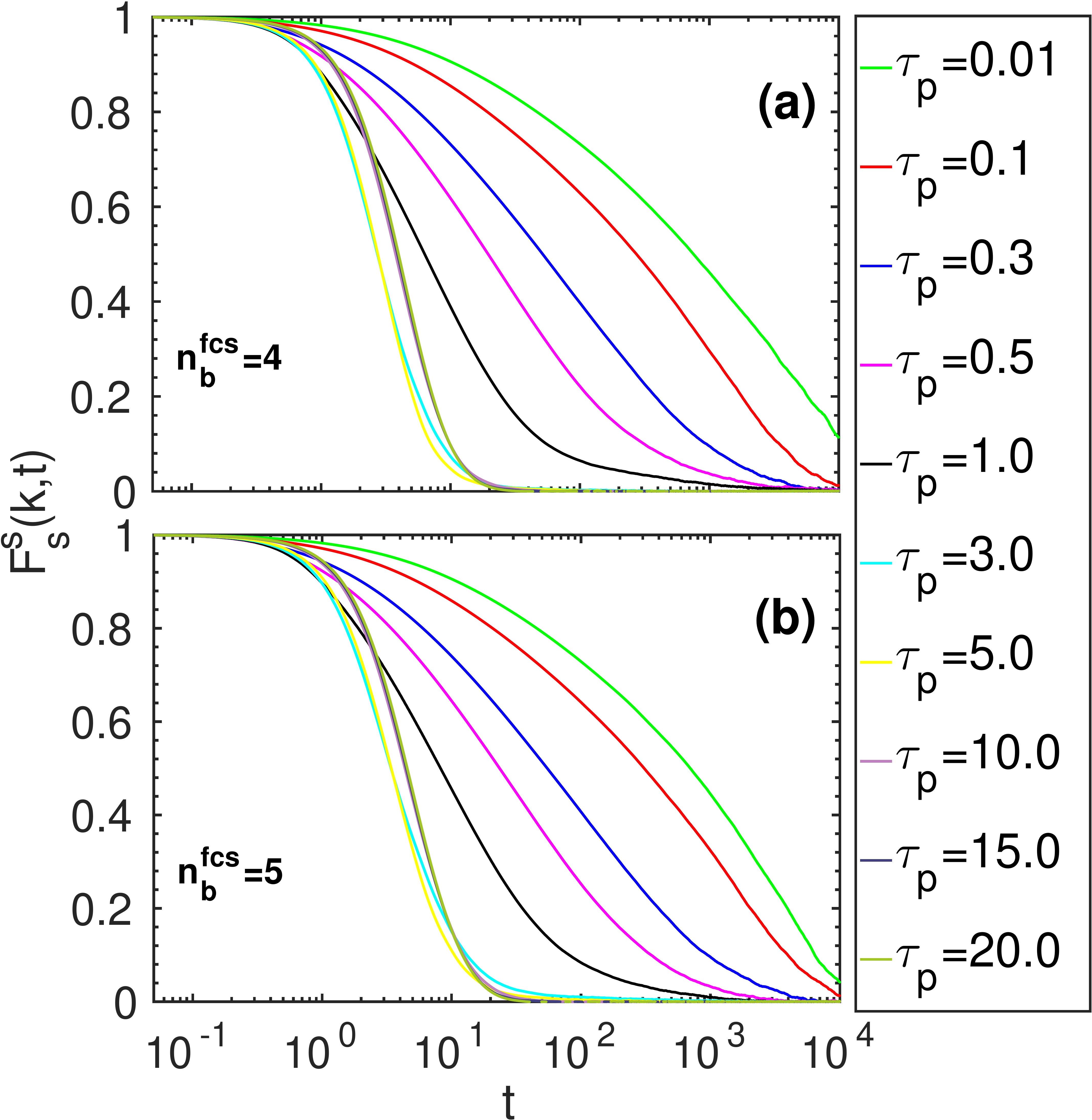}
	\caption{\label{f:fskts} Incoherent intermediate scattering function $F_s^s(k,t)$ of the 
	small particles at $f_a=$ 3.0 : (a) $n_b^{fcs}=$ 4 and (b) $n_b^{fcs}=$ 5. Relaxation
	of the $F_s^s(k,t)$ becomes faster with increment in the $\tau_p$.}
\end{figure}
\begin{figure}
	\includegraphics[width=8.5cm, height=4.0cm]{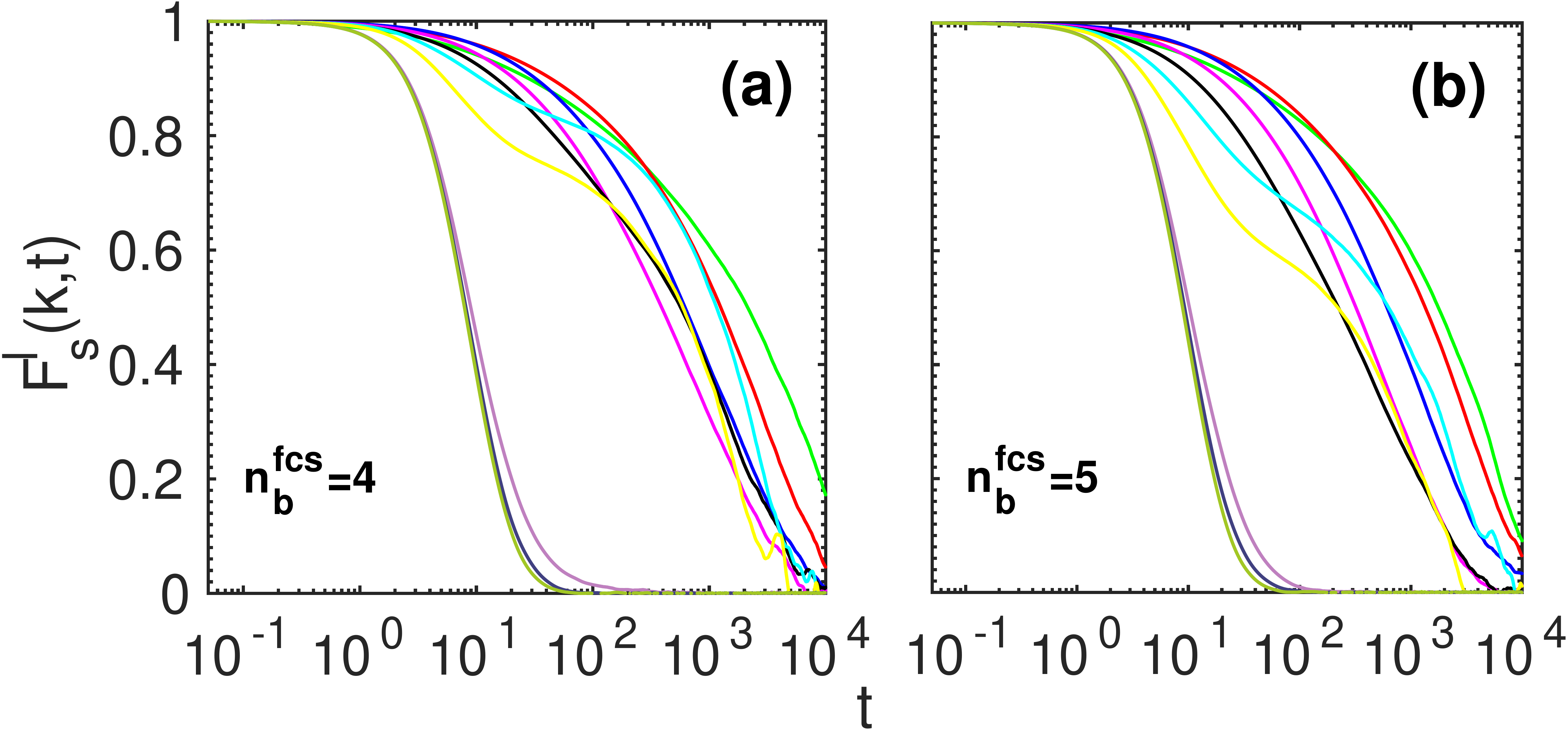}
	\caption{\label{f:fsktl} Incoherent intermediate scattering function $F_s^l(k,t)$ of the 
	large particles at $f_a=$ 3.0 : (a) $n_b^{fcs}=$ 4 and (b) $n_b^{fcs}=$ 5. Line colors 
	are according to the legend of Fig. \ref{f:fskts}. Relaxation of the $F_s^l(k,t)$ is
	non-monotonic with the $\tau_p$.}
\end{figure}

To characterize the relaxation dynamics of the active-passive mixture, we invoke 
incoherent intermediate scattering function. The dynamics of the system
is analyzed by computing the $F_s(k,t)$ of $A$ and $B$ particles at the wave numbers 
$k=$ 4.5 and 5.0, respectively; the definition of $F_s(k,t)$ is given in 
Eq. \ref{e:fskt}. Figures \ref{f:fskts} and \ref{f:fsktl} show $F_s(k,t)$ of
$B$ and $A$ particles, respectively, at the activity $f_a=$ 3.0 and $\tau_p=$ 0.01-20 for
the local neighbor threshold values $n_b^{fcs}=$ 4 and 5. 
On comparing $F_s(k,t)$ of passive $B$ particles [see Fig. \ref{f:stcdyn}(c)] with the
active $B$ particles (in Fig. \ref{f:fskts}), it is evident that the density relaxation
in the active-passive mixture becomes slower at the smaller $\tau_p$ because the 
$F_s(k,t)$ of passive $B$ particles relax to zero near $t\simeq5\times10^3$, whereas the
$F_s(k,t)$ of $B$ particles in active system is above zero even till the time $t=10^4$.
Interestingly, the $F_s(k,t)$ of $A$ and $B$ particles in the active system shows more slowing
down for $n_b^{fcs}=$ 4 compared to the $n_b^{fcs}=$ 5 at smaller $\tau_p$. This slowing down
is because the number density of active $B$ particles is more in case of $n_b^{fcs}=$ 4, 
showing that activity enhances the glassiness in the system at smaller $\tau_p$. At $f_a=$ 3.0, 
as the persistence time of active force increases (near the TTL), $F_s^s(k,t)$ shows the faster 
density relaxations similar to the liquids, which is due to the activity induced fluidization
in the system. Thus, our system is showing the non-monotonous characteristic of the active
dense systems at smaller $\tau_p$ (from its passive limit), as reported in a recent 
experimental \cite{j:januscolid_exp1} and simulation \cite{j:szamel_activ,j:szamel_activ_rev} 
studies of dense active colloidal systems.

At smaller $\tau_p$, $F_s^l(k,t)$ is qualitatively similar to the $F_s^s(k,t)$.
On the other hand, the density relaxations of the $A$ 
particles (shown in Fig. \ref{f:fsktl}) exhibit a contrasting behavior at the 
intermediate $\tau_p$, where the system phase separates into the hexatic-liquid and 
solid-liquid phases. The slowing down of the $F_s^l(k,t)$ at the intermediate $\tau_p$ is
because of the solidification that hinders the movement of the $A$ particles, enormously, and
the hump in the $F_s^l(k,t)$ becomes pronounced. At $\tau_p=$ 3.0 and 5.0, where the active
system shows the extent of the phase separation at $f_a=$ 3.0, it shows the 
glasslike density relaxations \cite{j:berthier}.
This is because all the $A$ particles are not in the phase separation region, though, few of 
the $A$ particles are mixed with the $B$ particles that are in the phase coexistence region.
The mixing of the $A$ and $B$ particles can be visualized from the configurations given in 
Fig. \ref{f:hexpop}(c). Thus, our structural analysis is supporting the dynamics of the 
two-dimensional active-passive binary mixture in its steady state.

\bibliography{qsps} 

\end{document}